\documentclass[journal,draftcls, onecolumn]{IEEEtran}
\usepackage{graphicx}
\usepackage{float}
\usepackage{amsmath}
\usepackage{amssymb}
\usepackage{bm}
\usepackage{cuted}
\usepackage{subfiles}
\usepackage{color}
\graphicspath{{../images/}{./images/}}

\usepackage{caption}
\usepackage{subcaption}

\usepackage{notoccite}
\usepackage{cite}

\usepackage{algorithm}
\usepackage{algorithmic}

\usepackage[utf8]{inputenc} %unicode support

\usepackage{amsthm}
\allowdisplaybreaks

\newcommand{\z}{\bm{z}}
\newcommand{\h}{\bm{h}}
\newcommand{\x}{\bm{\theta}}
\newcommand{\w}{\bm{w}}
\newcommand{\prob}{p}
\newcommand{\cov}{\bm{R}}

\newcommand{\atan}{\text{atan}}

\title{Wireless Localization for mmWave Networks in Urban Environments}

\author{\IEEEauthorblockN{Macey Ruble and \.{I}smail~G\"uven\c{c}} \\
\IEEEauthorblockA{Dept. Electrical and Computer Engineering, North Carolina State University, Raleigh, NC 27695}\\
Email: \{mcruble,iguvenc\}@ncsu.edu
\vspace{-0.7cm}
}

\begin{document}
\maketitle
% \begin{frontmatter}

% \begin{fmbox}
% \dochead{Research}
% \author[
%   addressref={aff1},               
%   email={mcruble@ncsu.edu, iguvenc@ncsu.edu}   
% ]{\inits{MR}\fnm{Macey} \snm{Ruble}}
% \author[
%   addressref={aff2},
%   email={iguvenc@ncsu.edu}
% ]{\inits{IG}\fnm{\.{I}smail} \snm{G\"uven\c{c}}}

% \address[id=aff1]{%                           % unique id
%   \orgname{Department of Electrical and Computer Engineering, North Carolina State University},                               
%   \city{Raleigh, NC},                              % city
%   \cny{US}                                    % country
% }
% \address[id=aff2]{%
%   \orgname{Department of Electrical and Computer Engineering, North Carolina State University},                               
%   \city{Raleigh, NC},                              % city
%   \cny{US}                                    % country
% }

% \begin{artnotes}
% %\note{Sample of title note}     % note to the article
% % \note[id=aff2]{Equal contributor} % note, connected to author
% \end{artnotes}

% \end{fmbox}% comment this for two column layout

% \begin{abstractbox}

\begin{abstract} % abstract
Millimeter wave (mmWave) technology is expected to be a major component of 5G wireless networks. Ultra-wide bandwidths of mmWave signals and the possibility of utilizing large number of antennas at the transmitter and the receiver allow accurate identification of multipath components in temporal and angular domains, making mmWave systems advantageous for localization applications.      
In this paper, we analyze the performance of a two-step mmWave localization approach that can utilize time-of-arrival, angle-of-arrival, and angle-of-departure from multiple nodes in an urban environment with both line-of-sight (LOS) and non-LOS (NLOS) links. Networks with/without radio-environmental mapping (REM) are considered, where a network with REM is able to localize nearby scatterers. Estimation of a UE location is challenging due to large numbers of local optima in the likelihood function.  To address this problem, a gradient-assisted particle filter (GAPF) estimator is proposed to accurately estimate a user equipment (UE) location as well as the locations of nearby scatterers. Monte Carlo simulations show that the GAPF estimator performance matches the Cramer-Rao bound (CRB).  The estimator is also used to create an REM.  It is seen that significant localization gains can be achieved by increasing beam directionality or by utilizing REM.
\end{abstract}


% \end{abstractbox}

% \end{frontmatter}

\section{Introduction}
The demand for wireless broadband communication has been growing rapidly, which has been the driving force for the emergence of 5G cellular networks.  It has recently been shown in the literature that millimeter wave (mmWave) technology is not only feasible for dynamic outdoor cellular networks, but can facilitate a thousand fold increase in data capacity \cite{zhu2014demystifying, rappaport2013millimeter, rappaport2015wideband, maccartney2015exploiting}.  The mmWave cellular networks are expected to first be deployed in \textit{dense urban environments} where the global positioning system (GPS) signal may typically be unavailable and the demand for large data rates is high.  With coverage ranges that can extend to hundreds of meters, the network must be solely responsible for localization, while also simultaneously achieving high data rates in such scenarios~\cite{zhu2014demystifying}. Communication performance in 5G networks will be limited by the amount of time required to align the highly directional beams of the communicating nodes, particularly for exhaustive beam searches, which are costly to capacity \cite{saloranta2017comparison}.  A network that is able to localize node positions can significantly reduce the time spent on beam alignment and increase capacity \cite{maschietti2017robust}. Thus, it is important to characterize the performance of mmWave network localization for urban scenarios. 

\begin{figure}[h!]
	\includegraphics[width=0.80\textwidth]{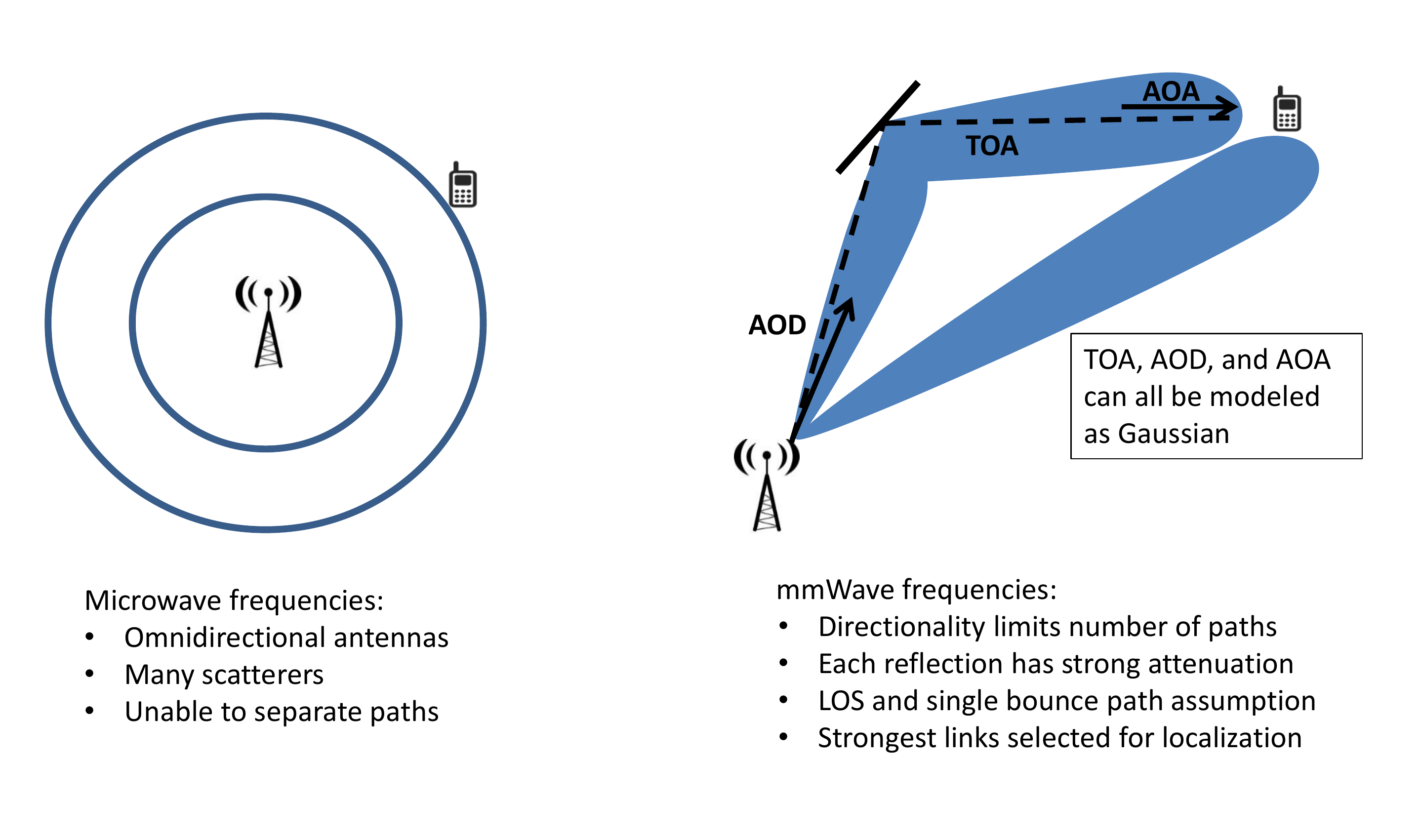}
	\centering
	\caption{Signal propagation in  microwave and mmWave frequencies.}
	\label{fig:mmWaveBsUwave}
\end{figure}

Wireless localization with strictly non-line-of-sight (NLOS) paths is achievable for omni-directional antennas by exploiting the time-of-arrival (TOA), angle-of-arrival (AOA) and angle-of-departure (AOD) measurements \cite{guvenc2009survey,miao2007positioning,han2016performance}.  While accurately measuring AOA and AOD is relatively difficult at lower frequencies due to the rich scattering  and poor path separability, this is much easier to achieve in mmWave channels leading to improved localization performance \cite{shahmansoori2017position}.  The mmWave frequencies also allow the use of ultra-wide bandwidths larger than 1 GHz, which helps in providing precise TOA estimates.  Fig. \ref{fig:mmWaveBsUwave} highlights channel characteristics of mmWave frequencies that offer advantages over traditional microwave frequencies for localization.  

% \textcolor{magenta}{IG: Introduction is quite long. To introduce a bit more structure, I would include a subsection title right here, saying ``Literature Review on mmWave Localization'', and another on contributions as noted below. I also made several updates in the red text below in introduction which I did not incorporate in the response document; you can update them.}

\subsection{Literature Review on mmWave Localization}

There have been  recent studies in the literature that evaluate mmWave localization performance in various scenarios.  Localization with received signal strength (RSS), TOA, and AOA are analyzed in \cite{lemic2016localization} (separately and jointly for different measurement parameters) for LOS and NLOS scenarios, showing promising results with TOA and AOA, and less reliable results with RSS.  A log-normal path loss model is used to evaluate RSS, time-difference-of-arrival (TDOA), and AOA localization methods for LOS paths in \cite{el2014evaluation}. A mobile's location and orientation are estimated
% \textcolor{magenta}{IG: maybe better to use ``estimated'' rather than ``considered'' here to be more specific (and considered is used in next sentence too, hence reword)}
jointly in \cite{guerra2015position, shahmansoori20155g} for mmWave systems.  It is shown in these papers that a single fixed equipment (FE) is sufficient to localize a user equipment (UE), but only LOS paths are considered.  A direct localization approach for a user connected with multiple FEs is introduced in \cite{garcia2017direct}, but NLOS paths are treated as interference and LOS paths are still required.

At lower frequencies NLOS paths are treated as interference.  A major advantage of mmWave frequencies is that very few paths have significant received signal strength, which results in channel sparseness \cite{deng2014mm}.  This property enables NLOS paths to no longer be considered as interference, but to instead be exploited as paths with additional information that can improve localization performance \cite{mendrzik2017harnessing}.  The work in \cite{shahmansoori2017position} shows that it is possible to use LOS and NLOS paths to determine the orientation and position of a node communicating with a single transmitter at mmWave frequencies under certain conditions. Specifically, it is shown that sufficient conditions for position and orientation estimation require at least one LOS path or three NLOS paths. The Cramer-Rao bound for localization and orientation is derived and a localization algorithm is proposed that exploits channel sparseness to estimate AOD/AOA/TOA, which is then used to estimate user position and orientation.  The work in \cite{abu2017error} extends the Cramer-Rao bound to three dimensions.

\subsection{Summary of the Proposed mmWave Localization Technique}

The algorithm proposed in \cite{shahmansoori2017position} is only suited for single transmitter scenarios.  However, there may be scenarios where a single transmitter is unable to establish enough paths to meet the sufficient conditions for localization. In this work, we consider scenarios that are not necessarily limited to a single transmitter.  Instead, an initial access or beam alignment stage is used to obtain rough AOD/AOA/TOA estimates for LOS and NLOS paths from one or multiple transmitters.  Then, the user position is estimated using the AOD/AOA/TOA estimates.  This step requires optimizing a non-convex cost function with many local maxima.  To accomplish this we propose a gradient-assisted particle filter (GAPF) estimator. While separately estimating TOA, AOD, and AOA is sub-optimal, it significantly reduces computational effort and allows links from multiple transmitting FEs to be used for localization. 

It should be noted that the work in \cite{shahmansoori2017position} studies methods to jointly estimate and refine AOD/AOA/TOA for LOS and NLOS paths from a single FE.  This process can be used to obtain improved AOD/AOA/TOA estimates for one FE at a time in place of the proposed reduced complexity first step.  Then, these estimates can be combined and used in the second step for improved localization accuracy over the proposed reduced complexity approach. The reduced complexity approach is only considered in this work because the minimal processing approach to estimating AOD/AOA/TOA greatly reduces computation time and still achieves adequate localization accuracy. The proposed method is best suited for environments where many nodes are connected, such as a 5G network with many FEs in an urban environment where a single FE may not be able to establish enough paths to meet the sufficient conditions for localization. 

Urban environments with many buildings and large structures contain scatterers that remain fixed in space.  The three dimensional spatial characteristics of the urban environment can be captured by radio-environmental mapping (REM) and later exploited to estimate scatterer locations for NLOS paths to improve localization \cite{landstrom2016transmitter}.  The knowledge provided by localization and REM can be used to relax initial access requirements and improve capacity for 5G communication systems \cite{witrisal2016high}.  In this work, we study REM-assisted localization, which assumes scatterer locations are estimated \emph{a~priori} and can be used to improve localization performance.

% \textcolor{magenta}{IG: Maybe another subsection here, titled ``Contributions of this Work''}

\subsection{Methods/Experiments and Contributions of This Work}

To our best knowledge, mmWave network localization that exploits LOS and NLOS paths from multiple FEs in an urban environment utilizing TOA/AOA/AOD with/without REM has not been studied in the literature.  This paper analyzes the localization performance of 5G mmWave communication networks specifically considering urban canyon and urban corner environments where directional beams are captured at the UE in the downlink from one or multiple FEs. The contributions of this paper are summarized as follows:
\begin{itemize}
    \item A reduced complexity two-step localization scheme is introduced where in the first step the receiver obtains rough estimates of AOD/AOA/TOA for LOS and NLOS paths from initial access or beam alignment stages from multiple FEs \cite{shahmansoori2017position},\cite{5GcellNet}. 
    % \textcolor{magenta}{IG: It would be good to add 1-2 references on beam alignment in previous sentence, especially if they talk about extracting such information.} 
    The AOD estimate is obtained from a transmitted reference signal where the transmitter embeds the center beam direction and the AOA/TOA are estimated non-jointly by the receiver.  The second step uses the AOD/AOA/TOA estimates to estimate the receiver and NLOS scatterer location coordinates.
    \item A gradient-assisted particle filter (GAPF) estimator is proposed as a maximum likelihood (ML) estimator to estimate the UE position and scatterer coordinates over a non-convex space. It is shown to have performance that matches the Cramer-Rao bound (CRB) through Monte-Carlo simulations.
    \item Localization performance is analyzed in urban canyon and urban corner scenarios where a UE is connected with one FE or two FEs.
    \item The localization accuracy of REM-assisted and non-REM-assisted network performance is analyzed. The performance of a \emph{perfect} REM system that has perfect knowledge of a scatterer locations (with no localization error) for each observed path is used to bound realistic REM systems.
    \item The scatterer locations that are extracted from the proposed localization approach are used to create an REM. 
\end{itemize}

The rest of this paper is organized as follows.  Section \ref{section:model} introduces the mmWave localization model and studies how LOS and NLOS paths can be separated for localization purposes. Section \ref{sec:estimation} discusses estimation methods and introduces the GAPF estimator for localization.  Section \ref{section:ExistenceLowerBounds} derives the CRB for mmWave localization performance. Section \ref{section:simulations} compares estimator performance to the CRB, analyzes mmWave localization performance in urban canyon/corner environments with/without REM, and examines the trade-off between system complexity and localization accuracy. Subsequently, the proposed estimator is used to create an REM.  Finally, Section \ref{sec:conclusion} provides concluding remarks.

\section{mmWave Localization System Model}\label{section:model}
This section introduces the  downlink system model in a mmWave network from multiple FE nodes to a single UE node, where the UE estimates its own position.  Without loss of generality, our results can also be extended to network-based localization relying on uplink mmWave signals from the UE. First, the 2D system model for mmWave urban localization is introduced.  Then, differentiating LOS from NLOS paths is studied.  Additionally, ray tracing techniques are used to further study the effects of directional beams at mmWave frequencies and justify that the results in the included 2D model are representative of a full 3D model.

\subsection{Localization Model}
While the position of a UE in a wireless network can be estimated directly, it is often more practical to implement a two-step positioning approach that first determines a set of parameters such as TOA, AOD, or AOA, which are then used to estimate the position \cite{gezici2008survey}.  

The considered scenario utilizes the periodic beam training stage or initial access for mmWave network communications to collect AOA/AOD/TOA for a variety of paths \cite{maschietti2017robust,andrews2016modeling}. The beam training stage searches over possible beams at the FE and UE where each directional beam points in a different direction. The beams with the strongest signals are then used to estimate AOA/AOD/TOA for the paths associated with those beams.
A propagation channel is known to be sparse at mmWave frequencies~\cite{deng2014mm,andrews2016modeling}, for which one major reason is the use of highly directional beams enabled at mmWave frequencies with many antenna elements.
%A mmWave frequency channel is sparse \cite{deng2014mm,andrews2016modeling}, which is a result of mmWave frequency antennas that have many elements and form highly directional beams. 
Hence, each beam will typically contain a LOS path or a strong single bounce NLOS path \cite{deng2014mm,rappaport2012cellular} along with multiple bounce NLOS paths that will have large attenuation and much lower signal strength \cite{andrews2016modeling}. Since only the strongest beams for each UE/FE pair are selected, multiple bounce paths are relatively unlikely, which allows the model to be simplified to LOS paths and single bounce NLOS paths \cite{andrews2016modeling}.

Assuming a synchronized network, ultra-wide bandwidths of mmWave frequencies provide precise TOA estimates \cite{koivisto2016joint}. Measuring AOA and AOD is not feasible at lower frequencies because of large numbers of scattered paths.  However, the arrays with large numbers of antenna elements at mmWave frequencies easily fit on chip enabling highly directional beams. Arrays at the UE provide precise AOA measurements, where it is assumed the orientation of the UE is known so that the AOA relative to the overall coordinate system is determined from the AOA relative to the UE array.  
An AOD measurement is obtained from the FE, which transmits the beam's \emph{quantized} AOD relative to the overall coordinate system.  This is easily calculated from the AOD of the FE relative to the antenna array under the assumption that the FE orientation is known.  Additionally, the FE is assumed to broadcast its position coordinates.
%An AOD measurement is obtained from the FE, which transmits the beam's \emph{quantized} AOD relative to the overall coordinate system.  This is easily calculated from the AOD of the FE relative to the antenna array under the assumption that the FE orientation is known.  Additionally, the FE is assumed to transmit it's position coordinates.  
These low complexity methods of obtaining AOA/AOD/TOA make mmWave an ideally suited technology for localization. 

\begin{figure}[t]
	\includegraphics[width=0.48\textwidth]{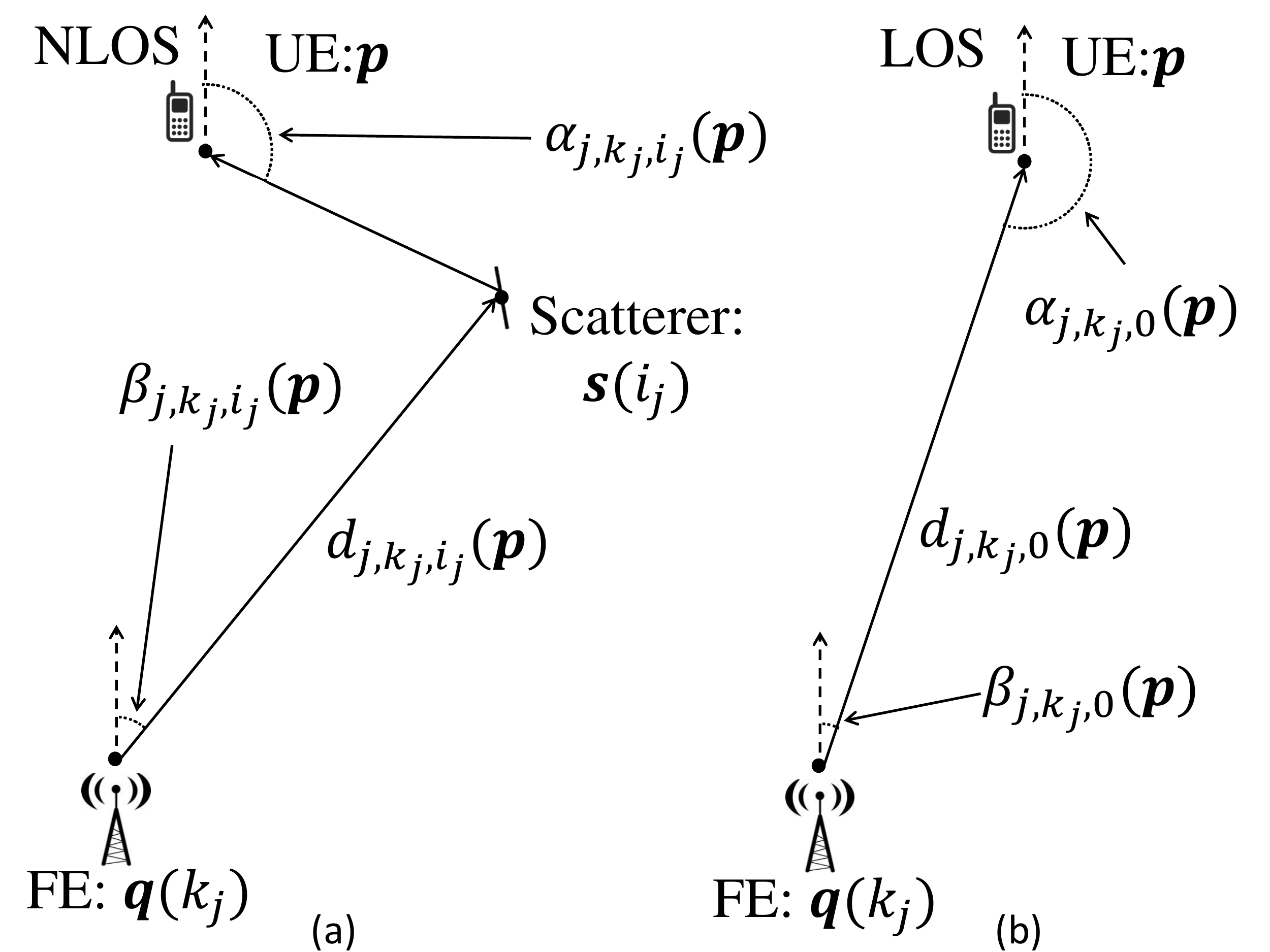}
	\centering
	\caption{mmWave localization model for (a) NLOS and (b) LOS scenarios.}
	\label{fig:model}
\end{figure}

From the beam training stage, we assume that the AOA/AOD/TOA are measured for paths corresponding with beams from $N_\text{FE}$ FEs.  Then, each path is identified as LOS or NLOS, which is discussed in \ref{sec:LOSvsNLOS}.  It is assumed there are $N_\text{L} \leq N_\text{FE}$ LOS paths and $N_\text{N}$ NLOS paths, which are used to estimate the location of a UE at position 
\begin{equation}\label{eq:UE}
    {\bm p}=[p_x ~ p_y]^T.
\end{equation}  
Fig.~\ref{fig:model} shows a NLOS and a LOS path along with each path's associated parameters, which are explained in the following paragraphs.

Let the vector $\bm{q} \in \mathbb{R}^{2 \times N_\text{FE}}$ be the 2-dimensional coordinate vector for $N_\text{FE}$ transmitting FEs and
\begin{equation} \label{eq:FE}
    \bm{q}(k_j)=[q_x(k_j)~q_y(k_j)]^T
\end{equation} 
be the location of FE $k_j$ that transmits path $j$.  The vector $\bm{s} \in \mathbb{R}^{2 \times N_\text{N}}$ is the 2-dimensional coordinate vector for the locations of all the scatterers of the NLOS paths, and 
\begin{equation}\label{eq:scatterer}
    \bm{s}(i_j)=[s_x(i_j)~s_y(i_j)]^T
\end{equation}
is the location of scatterer $i_j$ that reflects path $j$.  

Let $\bm{\alpha}(\x) \in \mathbb{R}^{(N_\text{L}+N_\text{N}) \times 1}$ contain the AOA for all LOS and NLOS paths, $\bm{\beta}(\x) \in \mathbb{R}^{(N_\text{L}+N_\text{N}) \times 1}$ contain the AOD for all LOS and NLOS paths, and $\bm{d}(\x) \in \mathbb{R}^{(N_\text{L}+N_\text{N}) \times 1}$ contain the total travelled path distance (which TOA will measure) for all LOS and NLOS paths.  A NLOS path $j$ transmitted from FE $k_j$ and reflected from scatterer $i_j$ will have AOA $\alpha_{j,k_j,i_j}$, AOD $\beta_{j,k_j,i_j}$, and distance $d_{j,k_j,i_j}$.  A LOS path $j$ transmitted from FE $k_j$ will have AOA $\alpha_{j,k_j,0}$, AOD $\beta_{j,k_j,0}$, and distance $d_{j,k_j,0}$ where the index $0$ signifies a LOS path.  

For consistency, let the first $N_\text{L}$ elements of $\bm{\alpha}(\x)$, $\bm{\beta}(\x)$, and $\bm{d}(\x)$ correspond to LOS paths so that these elements have indices $j=1,...,N_\text{L}$; $k_j=1,...,N_\text{L}$; and $i_j=0$. The remaining $N_\text{N}$ elements are the remaining NLOS paths with indices $j=N_\text{L}+1,...,N_\text{L}+N_\text{N}$; $k_j$ corresponds to the FE locations from which NLOS path $j$ originated; and $i_j=1,...,N_\text{N}$ corresponds to the scatterer location from which path $j$ reflects.  Then, similar to \cite{miao2007positioning}, the measured/observed parameters for LOS paths to be used for localization can be written as
\begin{align}
&\alpha_{j,k_j,0}(\x) = \text{atan2}\Bigg(\frac{q_x(k_j)-p_x}{q_y(k_j)-p_y}\Bigg), \label{eq:parameters1}\\
&\beta_{j,k_j,0}(\x) = \text{atan2}\Bigg(\frac{p_x-q_x(k_j)}{p_y-q_y(k_j)}\Bigg), \label{eq:parameters2}\\
&d_{j,k_j,0}(\x) = \sqrt{(p_x-q_x(k_j))^2 + (p_y-q_y(k_j))^2}. \label{eq:parameters3}
\end{align}
On the other hand, NLOS paths have the parameters
\begin{align}
&\alpha_{j,k_j,i_j}(\x) = \text{atan2}\Bigg(\frac{s_x(i_j)-p_x}{s_y(i_j)-p_y}\Bigg), \label{eq:parameters4}\\
&\beta_{j,k_j,i_j}(\x) = \text{atan2}\Bigg(\frac{s_x(i_j)-q_x(k_j)}{s_y(i_j)-q_y(k_j)}\Bigg), \label{eq:parameters5}
\end{align}
\begin{equation}
d_{j,k_j,i_j}(\x) = \sqrt{(s_x(i_j)-p_x)^2 + (s_y(i_j)-p_y)^2}% \nonumber\\ 
 + \sqrt{(s_x(i_j)-q_x(k_j))^2 + (s_y(i_j)-q_y(k_j))^2},\label{eq:parameters6}
\end{equation}
where atan2$(\cdot)$ is the four quadrant inverse tangent:
\begin{equation}\label{eq:atan2}
\atan2\bigg( \frac{y}{x} \bigg) = 
\begin{cases} 
      \atan\big(\frac{y}{x}\big) & \text{if}~ x \geq 0, \\
      \atan\big(\frac{y}{x}\big) + \pi & \text{if}~ x < 0 ~\text{and}~ y \geq 0, \\
      \atan\big(\frac{y}{x}\big) - \pi & \text{if}~ x < 0 ~\text{and}~ y < 0, \\ 
      \frac{\pi}{2} & \text{if}~ x=0 ~\text{and}~ y > 0, \\
      -\frac{\pi}{2} & \text{if}~ x=0 ~\text{and}~ y < 0, \\
      \text{undefined} & \text{if}~ x=0 ~\text{and}~ y=0,
   \end{cases}
\end{equation}
and $\atan(\cdot)$ is the ordinary arctangent function.

Our goal is to estimate $\bm{p} = [p_x~p_y]^T$ from noisy measurements of the parameters in \eqref{eq:parameters1}--\eqref{eq:parameters6} from multiple links with one or more FEs.  The noise of the AOA/AOD/TOA path parameters is assumed to be zero-mean Gaussian throughout this paper and the estimation algorithm is designed under the Gaussian noise assumption.  Though, the Gaussian distribution is just one possibility for parameter noise distributions.  The proposed algorithms can still be used if the parameters have non-Gaussian noise distributions, but with degraded performance. The localization estimator is optimal under the Gaussian noise scenario, but scenarios with non-Gaussian distributions are still expected to have reasonable performance \cite{kay1993statistical}. Nonetheless, the Gaussian assumption is reasonable for AOA/AOD/TOA at mmWave frequencies as supported by \cite{1611097,sahinoglu2008ultra,samimi20163}, where \cite{samimi20163} uses a measurement campaign to show that Gaussian noise is a good fit for AOD and AOA at mmWave frequencies in urban environments. TOA estimation can be used to determine the total distance in \eqref{eq:parameters3} and \eqref{eq:parameters6} traveled by a path for LOS and NLOS links, respectively.  The measured distance for a path is
\begin{equation}\label{eq:dprime}
    d_{j,k_j,i_j}' = d_{j,k_j,i_j}(\x) + n_{j,k_j,i_j}^{(d)},
\end{equation}
with $n_{j,k_j,i_j}^{(d)} \sim N(0,\sigma_{d_l}^2)$ where $l = \text{L}$ for LOS and $l=\text{N}$ for NLOS paths.  The center direction of the transmitted beam can be transmitted by the FE and used as an AOD estimate. The measured AOD is 
\begin{equation}\label{eq:betaprime}
    \beta_{j,k_j,i_j}' = \beta_{j,k_j,i_j}(\x) + n^{(\beta)}_{j,k_j,i_j},
\end{equation}
with $ n^{(\beta)}_{j,k_j,i_j} \sim N(0,\sigma_{\beta_l}^2)$ where $l = \text{L}$ for LOS and $l=\text{N}$ for NLOS paths.  The AOA can be measured by a receiver array so that the measured AOA is 
\begin{equation}\label{eq:alphaprime}
    \alpha_{j,k_j,i_j}' = \alpha_{j,k_j,i_j}(\x) + n^{(\alpha)}_{j,k_j,i_j},
\end{equation}
with $n^{(\alpha)}_{j,k_j,i_j} \sim N(0,\sigma_{\alpha_l}^2)$ where $l = \text{L}$ for LOS and $l=\text{N}$ for NLOS paths.  

Without loss of generality, the variance of the parameters have been assumed to only depend on whether the path is LOS or NLOS.  The parameters for the paths in \eqref{eq:parameters1}-\eqref{eq:parameters6} depend on the unknown position of the UE $\bm{p}$ and the unknown scattering locations $\bm{s}$, which are nuisance parameters. The vector of unknown parameters that must be estimated is
\begin{equation}\label{eq:paramSpace}
    \x=[\bm{p}^T~\bm{s}^T]^T\in \bm{\Theta},
\end{equation}
where $\bm{\Theta} \in \mathbb{R}^{2+2N_N}$ is the unknown parameter space.  We also consider networks that have REM capabilities.  REM provides information about the environment such as building locations and path link information, which can be used to determine scatterer locations and improve localization performance. The best case scenario is when the REM can provide sufficient information so that the scatterer locations are known perfectly.  Then, the estimator no longer has to estimate the scatterer locations for each NLOS path and only the UE position $\bm{p}$ needs to be estimated.  Therefore, the performance of REM aided localization is bounded by the case where scatterer locations are known and constant so that the UE position coordinates are the only parameters that need to be estimated and the unknown parameter vector becomes:
\begin{equation}\label{eq:paramSpaceREM}
    \x=[\bm{p}]\in \bm{\Theta},
\end{equation}
where $\bm{\Theta} \in \mathbb{R}^{2}$. With or without REM, the mmWave parameter measurements in \eqref{eq:parameters1}-\eqref{eq:parameters6} can be characterized using a nonlinear Gaussian model, which fits the general form:
\begin{equation}\label{eq:mmWaveMeas}
\z = \h(\x) + \w,
\end{equation}
where the nonlinear function $\h:\mathbb{R}^N \rightarrow \mathbb{R}^M$ and observations $\z$ are given by: 
\begin{equation*}
    \h(\x) = [\bm{\alpha}(\x)^T ~ \bm{\beta}(\x)^T ~ \bm{d}^T(\x)]^T ~~~\text{and}~~~ \z = [\bm{(\alpha}')^T ~ (\bm{\beta}')^T ~ (\bm{d}')^T]^T.
\end{equation*}
The measurement noise $\w \sim \mathcal{N}(0,\cov)$ is additive Gaussian noise from $n_{j,k_j,i_j}^{(d)}$, $n^{(\beta)}_{j,k_j,i_j}$, and $n^{(\alpha)}_{j,k_j,i_j}$ with measurement covariance matrix,
\begin{equation}\label{eq:measCovariance}
\cov = \text{diag}\Big( (\sigma_{\alpha_L}^2)_{1 \times \text{N}_L} , (\sigma_{\alpha_N}^2)_{1 \times \text{N}_N} , (\sigma_{\beta_L}^2)_{1 \times \text{N}_L} , (\sigma_{\beta_N}^2)_{1 \times \text{N}_N} , (\sigma_{d_L}^2)_{1 \times \text{N}_L},(\sigma_{d_N}^2)_{1 \times \text{N}_N} \Big).
\end{equation}
Localization then requires estimating the UE position $\bm{p}$ from the measurements $\z$.

\begin{table}[H] \scriptsize
	\begin{center}
	    \setlength{\tabcolsep}{0.7em}
	    {\renewcommand{\arraystretch}{1.4}
		\begin{tabular}{ |c|c|c|c|c|c|c| } 
			\hline
			~ & $\sigma_{{\alpha}_{\text{L}}} (^{\circ})$ &  $\sigma_{{\beta}_{\text{L}}} (^{\circ})$ & $\sigma_{{d}_\text{L}} (m)$ & $\sigma_{\alpha_{\text{N}}} (^{\circ})$ & $\sigma_{{\beta}_\text{N}} (^{\circ})$  & $\sigma_{d_\text{N}} ({m})$\\
			\hline
			28 GHz & $10.5$ & $8.5$ & $0.75$ & $10.1$ & $9.0$ & $0.75$ \\ 
			\hline
			73 GHz & $8.5$ & $ 5.5$ & $0.75$ & $6.0$ & $7.0$ & $0.75$ \\ 
			\hline
		\end{tabular}
		}
	\end{center}
\caption{Standard deviation of the observation parameters that are used in the simulations.}
\label{table:param}
\end{table}

\subsection{Statistics of AOA, AOD, and TOA}

The authors in \cite{samimi20163} use  measurements and simulations in urban environments to characterize the statistics of AOA, AOD, and TOA (needed for~\eqref{eq:measCovariance}) for a system with 2.5 ns multipath resolution and 800 MHz bandwidth, at center frequencies 28 GHz and 73 GHz. Horn antennas are used with $10.9^\circ$/$8.9^\circ$ azimuthal/elevation 3~dB beamwidths at 28 GHz and $7^{\circ}$/$7^\circ$ azimuthal/elevation 3~dB beamwidths at 73 GHz.  It has been shown that the parameters given closely capture the statistics of measurements in urban environments at these frequencies.  We use these parameters for each receiver in simulations, which are summarized in Table~\ref{table:param}.  Using the model from \cite{samimi20163} provides an accurate representation of an urban environment so that ray tracing is not necessary, which greatly simplifies simulations. It should be noted that the source of noise from the experimental campaign in \cite{samimi20163}, which leads to the variances in Table \ref{table:param} is from multipath in the same timing bin and the inability to separate them.  Antennas with more directionality will reduce the number of multipaths and reduce the noise variances from Table \ref{table:param} and improve localization accuracy.  The included AOD and AOA noise variances in Table~\ref{table:param} that are extracted from mmWave measurements in~\cite{samimi20163} are assumed to capture the effects such as NLOS reflection power loss, beamwidth, and interfering multipath components. 
%The included AOD and AOA noise variances in Table \ref{table:param} capture effects such as NLOS reflection power loss, beamwidth, and interfering multipath; and have been shown to accurately represent the mmWave frequency channel \cite{samimi20163}.}  
It should also be noted that these parameters assume the same receive SINR for both central frequencies of 28~GHz and 73~GHz as there are no interfering transmitters using the same bands; in other words, impact of central frequency is limited to the statistics reported on Table~\ref{table:param}, and interfering path behavior at these two different mmWave bands are not explicitly taken into account.  On the other hand, mmWave frequencies have little interference due to highly directional communications and highly attenuated reflections \cite{zhu2014demystifying}. 

\subsection{Differentiating LOS and NLOS Paths}\label{sec:LOSvsNLOS}

\begin{figure}[t]
	\centering
	\subcaptionbox{}{\includegraphics[width=0.18\textwidth]{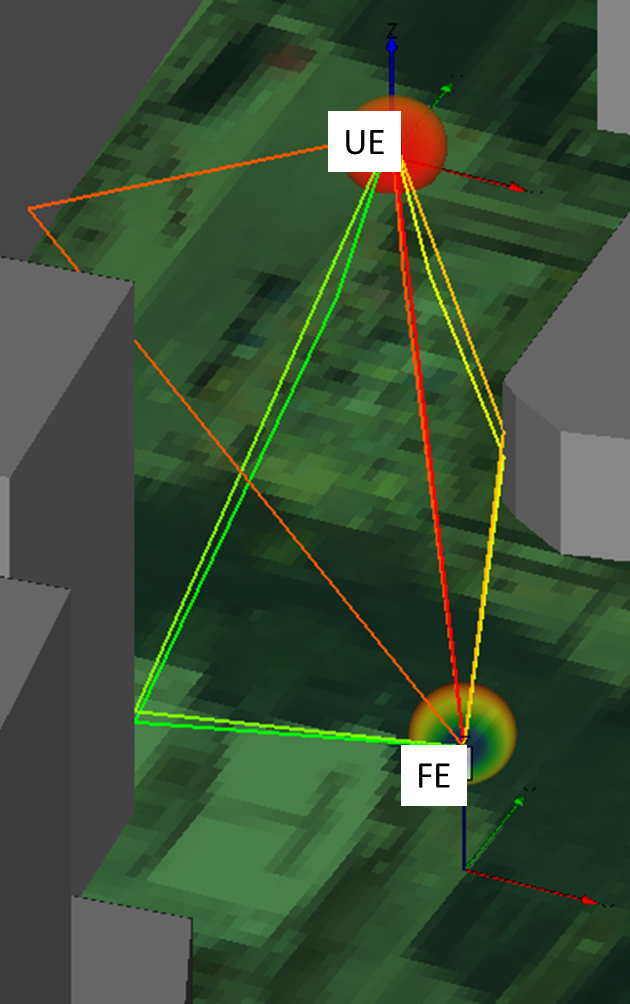}}%
	%\subcaptionbox{}{\includegraphics[width=0.26\textwidth]{CRLB}}%
	\subcaptionbox{}{\includegraphics[width=0.41\textwidth]{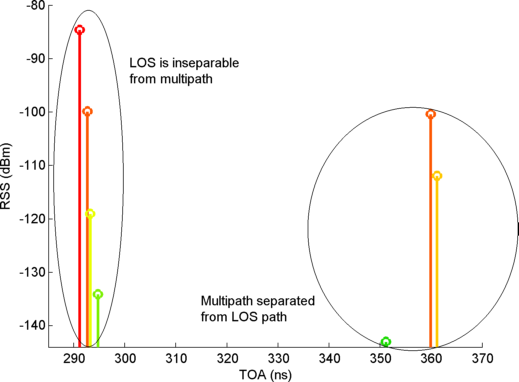}}%
	\subcaptionbox{}{\includegraphics[width=0.41\textwidth]{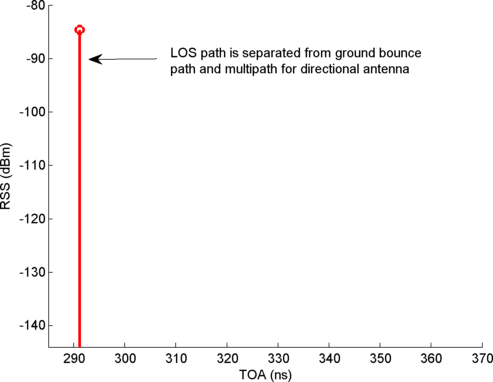}}%
	\caption{(a) Geometry and ray tracing paths for Wireless Insite simulation in an urban environment.  Paths are colored based on RSS where the strongest RSS path is red and the weakest RSS paths is green. (b) TOA of simulation with $80^\circ$ 3 dB beamwidth. (c) TOA of simulation with $28^\circ$ 3 dB beamwidth.}
	\label{fig:insiteSim}
\end{figure}

A challenging aspect for mmWave localization will be LOS/NLOS path identification and path separation, which is needed for generating the vector of unknown parameters in~\eqref{eq:paramSpace} (includes one unknown scatterer location for each NLOS path, and no scatterer for a LOS path),  the covariance matrix in~\eqref{eq:measCovariance}, and hence the likelihood function to be defined in~\eqref{eq:likelihood}. 
To this end, beam directionality can play an important role in LOS/NLOS differentiation and path separation, since it significantly impacts the multipath characteristics in mmWave systems. To gain further insights, an example with Remcom's Wireless Insite mmWave ray tracing simulator is used to simulate path separation with different directionalities of beams.  A center frequency of 73 GHz is used and a geometry is chosen that has multipath from ground bounce and building reflection as seen in Fig.~\ref{fig:insiteSim}(a). The FE is at a height of $10$ m and the UE is at a height of $2$ m.  The geometry is held fixed and two different beamwidths are used.  The first beam is semi-directional with a $80^\circ$ $3$ dB beamwidth and the second is a directional beam with a $28^\circ$ $3$ dB beamwidth.
Fig.~\ref{fig:insiteSim}(b) shows the TOA for the $80^\circ$ 3 dB beamwidth, where two clusters are observed, one of which has the LOS path along with multiple NLOS paths including ground bounce and reflected building paths.  In this case the LOS path is inseparable from others as mmWave systems will have a typical multipath resolution of $2.5$ ns.  This leads to a noisy AOA measurement and reduced accuracy.  
On the other hand, Fig. \ref{fig:insiteSim}(c) shows the TOA for the beam with the $28^\circ$ 3~dB beamwidth.  In this case, the directional beam gain causes all multipath including the ground bounce path to be negligible and only the LOS path is detected.  Since the 3D analysis of directional beams are able to separate out ground bounce and multipath from the main path, for simplicity it is reasonable to use a 2D model as path separation and localization results will be similar to a 3D model.

Directional beams and unique properties of mmWave such as signal attenuation, small RMS delay spread, and sparsity make differentiating individual paths and separating LOS path from NLOS paths feasible~\cite{shahmansoori2017position,rappaport2012cellular}.  The work in \cite{shahmansoori2017position} uses sparse estimation to separate and identify paths, but with added computational complexity. Simpler methods also exist that can separate LOS from NLOS paths. For example, the AOD and AOA can be compared using the overall coordinate system, which can be computed from the AOD and AOA relative to the antenna arrays if the FE and UE orientation are known. Then, a LOS path will have the parameters \eqref{eq:parameters1}-\eqref{eq:parameters6} with an AOD $\beta$ and AOA $\alpha$ such that 
%\begin{equation}
    $|\alpha-\beta| = \pi$. 
%\end{equation}
However, noisy versions of the parameters are observed with measured AOD $\beta'$ and measured AOA $\alpha'$.  Therefore, a threshold $\xi$ can be introduced that identifies the paths as LOS or NLOS for a desired probability of error:
\begin{equation}\label{eq:Anglethresh}
I_{\text{LOS}}(\alpha',\beta') = 
\begin{cases}
    1, & \text{if}~ |\alpha'-\beta'|-\pi \leq \xi \\
    0, & \text{if}~  |\alpha'-\beta'|-\pi > \xi
\end{cases}~,
\end{equation}
where $I_{\text{LOS}}(\alpha',\beta')$ is an indicator function, which is $1$ if LOS, and $0$ otherwise.
Another approach to separate LOS and NLOS paths is to exploit TOA and RSS.  Urban scenario path loss for 5G systems is studied in \cite{sun2016propagation} and \cite{rappaport2013millimeter} where it is shown that LOS and NLOS paths have different RSS statistics and a reflection from a NLOS path will result in $4$-$6$ dB of power loss.  Therefore, the path distance from a TOA estimate can be compared to the theoretical RSS path loss in free space to differentiate between LOS and NLOS paths.

\section{mmWave Location Estimation}\label{sec:estimation}

In this section, we will first describe the likelihood function for a UE's location based on AOA, AOD, and TOA observations. Subsequently, we will study gradient methods, particle filter methods, and a gradient-assisted particle filter technique to solve the two-step mmWave localization problem.  

\subsection{Maximizing the Likelihood Function}
It is assumed that the distributions of the parameter observations \eqref{eq:dprime}-\eqref{eq:alphaprime} are known and used to calculate the covariance $\cov$ as in \eqref{eq:measCovariance}.  Then, an estimator for an unknown UE location can be obtained by maximizing the likelihood of a measurement \eqref{eq:mmWaveMeas} as follows:
\begin{equation}\label{eq:maxLikelihood}
\hat{\x} = \arg\max\limits_{\x} ~ \prob(\z;\x),
\end{equation}
where the likelihood for a measurement is
\begin{equation}\label{eq:likelihood}
\prob(\z;\x) = \frac{1}{\sqrt{(2\pi)^M|\cov|}} \exp\Bigg\{-\frac{1}{2}\bigg[\z-\h(\x)\bigg]_\mathcal{A}^T \cov^{-1} \bigg[\z-\h(\x)\bigg]_\mathcal{A}\Bigg\},
\end{equation}
and 
\begin{equation}
    [\z-\h(\x)]_\mathcal{A} = 
    \begin{bmatrix}
    \bm{m}(\bm{\alpha}'-\bm{\alpha})^T & \bm{m}(\bm{\beta}'-\bm{\beta})^T & (\bm{d}'-\bm{d})^T
    \end{bmatrix}
\end{equation}
enforces the difference between angular parameters (AOD and AOA) to be in the range $[-\pi,\pi]$.  This is achieved with the following modulus function that forces its argument into the interval $[-\pi,\pi]$:
\begin{equation}
    \bm{m}({\bm{x}}) = \bm{x} - 2\pi \bigg\lfloor \frac{\bm{x}+\pi}{2\pi} \bigg\rfloor.
\end{equation}
This then prevents a linear treatment of angles, which have a circular coordinate system and two angles cannot have a magnitude difference greater than $\pi$.  Without the modulus, the likelihood function will have discontinuities that result from the $\atan2$ function in \eqref{eq:parameters1}, \eqref{eq:parameters2}, \eqref{eq:parameters4}, and \eqref{eq:parameters5}, which results in estimators with inherent bias \cite{bashan2007estimation}.

\subsection{Non-REM Assisted Localization}
Networks without REM must estimate the UE coordinates $\bm{p}$ jointly with the scatterer locations (nuisance parameter) $\bm{s}$.  Thus, maximization of the likelihood function as in \eqref{eq:maxLikelihood} is required over the $2 + 2N_N$ dimensions of $\x=[\bm{p}^T~\bm{s}^T]^T$ to localize a UE.  The nonlinearity of \eqref{eq:mmWaveMeas} requires a nonlinear estimation algorithm to maximize \eqref{eq:likelihood}. Furthermore, localization without REM is particularly challenging as it leads to a non-convex likelihood function that must be maximized over many dimensions.  The global maximum of the likelihood function is the optimal estimate. 

\subsection{REM Assisted Localization}
The REM provides a map of the estimated building and scatterer locations, which can be used to estimate a scatterer location for an NLOS path.  It should be noted that a realistic implementation of the REM requires addressing many challenging aspects.  A major challenge in a realistic REM implementation is linking a measured NLOS path's AOA/AOD/TOA parameters to a particular scatterer location.  One method of achieving REM is to use AOA/AOD/TOA to compute and store the scatterer location for each NLOS path received by a node in a communicating network.  This allows a database to be created that serves as an REM map, which stores scatterer locations as well as AOA/AOD/TOA and other path information for each observed path.  Then, the scatterer location from an unknown NLOS path can be linked with an observed AOA/AOD/TOA by searching the REM map database for scatterers that are associated with similar path parameters. The authors in \cite{landstrom2016transmitter} and \cite{witrisal2016high} provide algorithms that use REM to reduce the search space for UE path tracking. Further details %are beyond the scope of this paper and 
on REM implementation is left as future work. 

Instead of focusing on REM implementation, we use a \emph{perfect} REM system with zero scatterer location estimation error to provide bounds on the limits of REM performance.
This serves as a performance bound to any realistic REM system that can be implemented. Thus, the performance of a real REM system will be in between the perfect REM bound and the bound of a system without REM.  A perfect REM system has knowledge of scatterer locations $\bm{s}$ so that $\bm{s}$ can be treated as a constant and the dimensionality of the likelihood function is reduced to $2$ dimensions so that a UE is localized by maximizing the likelihood function over $\x=[\bm{p}]$.

\subsection{Gradient Methods}
Gradient optimization methods such as gradient descent, Newton's method, Gauss-Newton method, and Levenberg-Marquardt method are often used for maximizing likelihood functions.  These algorithms all rely on the first and second derivatives of the nonlinear function to iteratively find the maximum of \eqref{eq:likelihood}.  However, these optimization methods alone are not guaranteed to find the global maximum or even converge.  If \eqref{eq:likelihood} has many local maxima, these methods will converge to the local maximum that is closest to the initial estimate.  An advantage of these methods is that they are computationally efficient.  The function \textit{\textit{lsqnon}lin} in MATLAB implements the ``trust-region-reflective'' algorithm, which is a variant of Newton's method.  It will be referred to as ``gradient method'' for the rest of this paper as any of the other mentioned gradient-based algorithms will give similar performance.  

\subsection{Particle Filters}
As an alternative to gradient methods, an exhaustive search estimator is guaranteed to find the global maximum, but is not computationally feasible for maximization over high dimensional functions.  Particle filters can be used for multivariate nonlinear estimation and attempt to approach the performance of exhaustive searches with less computational effort \cite{closas2006particle}.  Each particle corresponds to a point where the likelihood function is evaluated.  Instead of computing the cost function over the entire space, particle filters use random particle searches to focus on likelihood maxima. However, the number of particles required to ensure that the global maximum is reached increases with search space dimensionality.  The sequential importance resampling (SIR) particle filter \cite{ristic2004beyond} is a simple technique compared to other types of particle filters and is modified in our implementation.

SIR particle filters are typically used to track a target state vector $\x_k\in\mathbb{R}^N$ in a nonlinear time-dependent system where $t = t_k$ is the time instant at sample index $k \in \mathbb{N}$.  The target state evolves in time by
\begin{equation}
\x_k = \bm{f}_{k-1}(\x_{k-1}) + \bm{v}_{k-1},
\end{equation}
where $\bm{f}_{k-1}(\cdot)$ is a nonlinear function from the target model and $\bm{v}_{k-1}$ is process noise, which is included for errors in the model.  A measurement of the target state at time $t=t_k$ is given by:
\begin{equation}\label{eq:PFmeas}
\z_k = \h_k(\x_k) + \w_k,
\end{equation}
where $\h_k$ is a nonlinear measurement function and $\w_k$ is measurement noise.  Typically, a particle filter is used to estimate the target state $\x_k$ as it evolves in time (as opposed to $\h(\x)$ in \eqref{eq:mmWaveMeas}) from the mesurements $\z_k$.  However, it can also be used as a maximum-likelihood estimator.  Instead of letting $k$ represent a state in time, let it represent an iteration and the target not evolve; in other words, let the observations be given by:
\begin{equation}\label{eq:PFtargetState}
\x_k = \x_{k-1} + \bm{v}_k.
\end{equation}
Then, instead of estimating an evolving target state at each instant of time with an additional measurement $\z_k$, the target state estimate is improved with each iteration $k$ from a single measurement from \eqref{eq:mmWaveMeas}.  The target state $\x_k$ then represents a collection of $N$ particles, which are evaluations of the likelihood function where $\x_k^i$ is used to represent an individual particle.  Under these modifications, the measurement and process noise from \eqref{eq:PFmeas} and  \eqref{eq:PFtargetState} are Gaussian, $\bm{w}_{k} \sim \mathcal{N}(0,\cov_k)$ and $\bm{v}_{k-1} \sim \mathcal{N}(0,\bm{Q}_{k-1})$ so that,
\begin{align}
\prob(\z_k|\x_k) &= \mathcal{N}(\h(\x_k),\cov_k), \label{eq:PFlikelihood} \\
\prob(\x_k|\x_{k-1}) &= \mathcal{N}(0,\bm{Q}_{k-1}). \label{eq:PFspread}
\end{align}  
Since each iteration uses the same measurement ($\z_k = \z$), it is seen that \eqref{eq:PFlikelihood} is identical to \eqref{eq:likelihood} if the covariance $\cov_k=\cov$ and state $\x_k=\x$.  

\subsection{Gradient-Assisted Particle Filter}

\begin{algorithm}[t]
\caption{GAPF Iteration: $\{ \x_k^i \}_{i=1}^N = \text{GAPFiteration}(\{ \x_{k-1}^{i},w_{k-1}^i \}_{i=1}^{N},\z_k)$ }
\begin{algorithmic}\label{alg:PF}
\STATE $\bullet$  Resample the particles: $\{ \x_k^i \}_{i=1}^{N} = \text{RESAMPLE}\big(\{ \x_{k-1}^i,w_{k-1}^i \}_{i=1}^N\big)$
\FOR {$i = 1:N$}
\STATE $\bullet$  Draw particles $\x_k^i \sim \prob(\x_k|\x_{k-1}^i)$
\STATE $\bullet$  Gradient method estimation with each particle $\{ \x_k^i \}_{i=1}^{N}=\text{GM}\big(\{ \x_k^i \}_{i=1}^{N}\big)$
\STATE $\bullet$  Evaluate likelihood function $\tilde{w}_k^i = \prob(\z_k|\x_k^i)$
\ENDFOR
\STATE $\bullet$  Sum likelihood function evaluations: $t = \Sigma_{i=1}^{N} \tilde{w}_k^i$
\FOR {$i = 1:N$}
\STATE $\bullet$  Normalize: $w_k^i = \tilde{w}_k^i/t$
\ENDFOR
\STATE $\bullet$  State estimate is particle with largest weight from all iterations: $\hat{\x} = \x_l^j:~ w_l^j = \max\big(\{w_m^i\}_{m=1}^{k}\big)$
\end{algorithmic}
\end{algorithm} 

The proposed GAPF estimation algorithm is a joint particle filter gradient method algorithm that is able to overcome the weaknesses of the individual gradient based and particle filter based algorithms.  The gradient method is used to find likelihood function maxima nearest to particles, which reduces the number of particles required to search over the parameter space.  The particle filter enables random searches of the parameter space in search of other maxima, which aims to eliminate convergence to local maxima.

\begin{figure}[t]
	\includegraphics[width=0.80\textwidth]{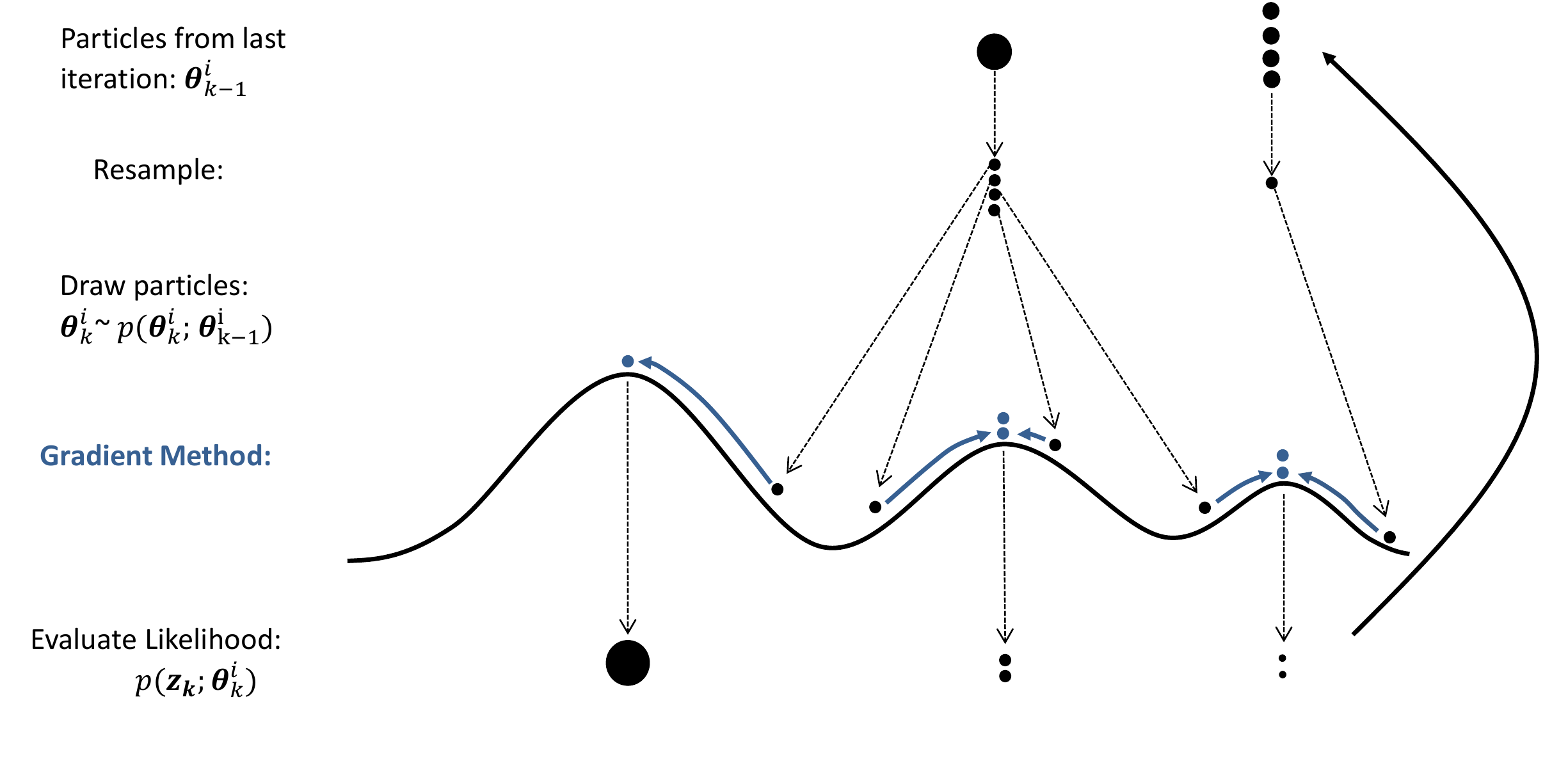}
	\centering
	\caption{One iteration of the GAPF estimator.}
	\label{fig:PFgraphic}
\end{figure}

Each iteration of the GAPF estimator in algorithmic form \cite{ristic2004beyond} is given in Algorithm~\ref{alg:PF}.  First, the particles are resampled using the particles $\x_{k-1}^{i}$ and weights $w_{k-1}^i$  from the last iteration.  The resampling function is taken from \cite{ristic2004beyond}, which draws particles in proportion to the weights so that particles that correspond to small likelihood evaluations are replaced by duplicate particles with larger likelihood function evaluations.  Then, the particles are drawn according to $\x_k^i \sim \prob(\x_k|\x_{k-1}^i)$ from \eqref{eq:PFspread}, which spreads out the particles randomly with the process noise covariance $\bm{Q}_{k-1}$ defined by the user.  Subsequently, the gradient method is applied to these drawn particles to find the nearest local maxima.  The likelihood of each particle is evaluated according to \eqref{eq:PFlikelihood} to determine the weight for that iteration.  The particle with the greatest weight is the target parameter estimate.  An iteration of the particle filter estimator is visualized in Fig.~\ref{fig:PFgraphic}.

Each iteration of the particle filter begins on the peaks of a set of local maxima.  The particles are randomly spread out to surrounding maxima, and the gradient method finds the peaks of the surrounding maxima.  Typically, mmWave localization leads to a likelihood function with a cluster of peaks.  The estimator must iterate until all of the peaks have been located to find the global maximum.  

\subsection{Initialization}
An initial grid search is required for estimation where the grid points/particles for the initial search $\x^i_0$ are fed into the first iteration of the estimator from Algorithm \ref{alg:Initialization}.  The grid must be extensive enough to guarantee that at least one of the initial particles in the grid is close to the likelihood function maximum, but not so extensive that it adds unnecessary computation.  Sufficient conditions for localization require a minimum of one LOS path, or two NLOS paths.  It is noted that \cite{shahmansoori2017position} requires three NLOS paths.  The discretion occurs because we assume the UE orientation is known and \cite{shahmansoori2017position} additionally estimates orientation.

Each parameter being estimated adds another dimension that the initial grid must cover.  Scenarios with only LOS paths only require a 2D grid to estimate the UE coordinates.  On the other hand, NLOS scenarios require the 2D UE coordinates to be estimated in addition to scatterer coordinates for each NLOS path. This adds an extra two dimensions that must be searched over for each NLOS path, making it difficult to cover the entire search space.
One option for scenarios with many NLOS paths is to jointly use all paths, which becomes computationally expensive as it requires many particles to cover the high dimensionality that results from all of the unknown scatterer locations.  A second option takes advantage of the fact that only two NLOS paths are required to localize the UE and estimate both scatterers.  Then, two or three NLOS paths can be separated from the other paths and used to estimate the UE location and scatterer locations for those paths.  This can be done for different sets of paths until each path has been used, which results in scatterer location estimates for all paths with significantly less numbers of required overall particles.  Then, the UE location estimates from every set of paths can be averaged to give a better estimate for the UE location.

Instead of a full exhaustive grid search in each dimension for the initial particles $\x^i_0$, an alternative initialization is proposed as seen in Algorithm \ref{alg:Initialization} that performs a separate grid search that significantly reduces the number of grid points required for scenarios with many NLOS paths.  Algorithm \ref{alg:Initialization} reduces the dimensionality of the search space by selecting paths such that a grid search is performed with a series of 2D grids rather than the entire grid space at once.  For example, a scenario with two LOS paths and two NLOS paths is initialized by first only using the two LOS paths to search over the 2D UE coordinates for an initial estimate.  Then, the estimated UE coordinates are held fixed and each of the NLOS paths are analyzed individually with a 2D grid search for the scatterer location.  It should be noted that NLOS scenarios require the use of at least two paths in order to have enough measurements to calculate the UE coordinates, which also require estimating the scatterer locations.  A single grid point is generated from initialization, which is fed into the particle filter as $\x_0$.  The initial grid point gives a rough estimate of the peak and the GAPF estimator searches around this point to find a better estimate.  
\begin{algorithm}[t]
\caption{Grid Initialization: $N_L$ LOS paths and $N_N$ NLOS paths}
\begin{algorithmic}\label{alg:Initialization}
\IF {$N_L > 0$}
\STATE $\bullet$ Search over UE coordinates ($p_x,p_y$) using $N_L$ LOS paths as measurements.  \\
$\bullet$ Let the coordinates that maximize the likelihood be ($\hat{p}_x,\hat{p}_y$).
\IF {$N_N > 0$}
\FOR{j=1:$N_N$} 
\STATE $\bullet$ Hold UE coordinates fixed at ($\hat{p}_x,\hat{p}_y$) and search over scatterer $j$ coordinates ($s_x(j),s_y(j)$) using all LOS paths and NLOS path $j$ as measurements. \\
$\bullet$ Let the coordinates for scatterer $j$ that maximize the likelihood be ($\hat{s}_x(j),\hat{s}_y(j)$).
\ENDFOR
\ENDIF
\ELSE 
\STATE $\bullet$ Search over UE coordinates ($p_x,p_y$) and scatterer coordinates ($s_x(m),s_y(m),s_x(n),s_y(n)$) using NLOS paths $m$ and $n$ as measurements. \\
$\bullet$ Let the grid point that maximizes the likelihood have UE coordinates ($\hat{p}_x,\hat{p}_y$) and scatterer coordinates $\hat{s}=(\hat{s}_x(m),\hat{s}_y(m),\hat{s}_x(n),\hat{s}_y(n))$.
\FOR{j=1:$N_N$ ($j \neq m,n$)} 
\STATE $\bullet$ Hold UE coordinates fixed at ($\hat{p}_x,\hat{p}_y$) with other scatters fixed at $\hat{s}$ and search over scatterer $j$ coordinates ($s_x(j),s_y(j)$) using NLOS path $j$ as measurements.  \\
$\bullet$ Let the coordinates for scatterer $j$ that maximize the likelihood be ($\hat{s}_x(j),\hat{s}_y(j)$) and add to $\hat{s}$.
\ENDFOR
\ENDIF
\STATE $\bullet$ Initial estimate is $\x_0=\{\hat{p}_x,\hat{p}_y,\hat{s}_x(i=1:N_N),\hat{s}_y(i=1:N_N)\}$.
\end{algorithmic}
\end{algorithm} 

\section{Fundamental Lower Bounds for mmWave Localization} \label{section:ExistenceLowerBounds}
Lower bounds on mmWave localization performance can provide insight into the limits of 5G networks and are helpful in identifying key factors when designing a network to meet certain specifications.  The CRB is often used to bound localization performance as it provides a lower bound on the covariance $\bm{C}_{\hat{\x}}$ of any \emph{unbiased} estimator $(\hat{\x})$ that satisfies the regularity conditions in \cite{kay1993statistical}: 
\begin{equation}\label{eq:CRLB}
\bm{C}_{\hat{\x}} - \bm{I}^{-1}(\x) \geq \bm{0} ,
\end{equation}
where $\bm{I}(\x)$ is the Fisher information matrix and $\geq \bm{0}$ represents a positive semidefinite matrix.  The covariance on estimator $\hat{\x}$ is defined as,
\begin{equation}\label{eq:covariance}
    \bm{C}_{\hat{\x}} = E\bigg[ \bigg(\hat{\x}-E[\hat{\x}]\bigg)\bigg(\hat{\x}-E[\hat{\x}]\bigg)^T \bigg]. 
\end{equation}
Additionally, using \eqref{eq:CRLB} it can be shown that the variance for element $m$ of $\hat{\x}$ is bounded by,
\begin{equation}\label{eq:boundElem}
    \text{var}(\hat{\x}_m) = [\bm{C}_{\hat{\x}}]_{mm} \geq [\bm{I}^{-1}(\x)]_{mm},
\end{equation}
for all $m$.  Since the observations are assumed Gaussian $\z \sim \mathcal{N}(\h(\x),\cov)$ and the covariance does not depend on $\x$, each element of Fisher information matrix is given by \cite{kay1993statistical}:
\begin{equation}\label{eq:GaussianFisher}
[\bm{I}(\x)]_{m,n} = \Bigg[ \frac{\partial \h(\x)}{\partial \x_m} \Bigg]^T \cov^{-1} \Bigg[ \frac{\partial \h(\x)}{\partial \x_n} \Bigg],
\end{equation}
where $1 \leq m,n \leq M$ and $\frac{\partial \h(\x)}{\partial \x_m}$ is the $m^{\text{th}}$ column of the Jacobian:
\begin{equation} \label{eq:Jacobian}
    \frac{\partial \h(\x)}{\partial \x} =
    \begin{bmatrix}
    \frac{\partial h_1(\x)}{\partial \theta_1} & \frac{\partial h_1(\x)}{\partial \theta_2} & \cdots & \frac{\partial h_1(\x)}{\partial \theta_N} \\
    \frac{\partial h_2(\x)}{\partial \theta_1} & \ddots & & \\
    \vdots & & & \\
    \frac{\partial h_M(\x)}{\partial \theta_1} & \cdots & &\frac{\partial h_M(\x)}{\partial \theta_N}
    \end{bmatrix}.
\end{equation}
% \begin{equation}
% \frac{\partial \h(\x)}{\partial \x_m} = \Bigg[ \frac{\partial [\h(\x)]_1}{\partial \x_m} ~ \frac{\partial [\h(\x)]_2}{\partial \x_m} ~ \cdots ~ \frac{\partial [\h(\x)]_M}{\partial \x_m} \Bigg]^T.
% \end{equation}
It should be noted that the estimator only approaches the performance bound if the observation noise is actually Gaussian.  However, the CRB still bounds non-Gaussian scenarios as other noise distributions will lead to degraded performance \cite{kay1993statistical}.

For completeness, the individual elements of the Jacobian matrix are calculated in Appendix~\ref{Sec:appendixa}.  For an unbiased estimator, we need to have $E[\hat{\x}]=\x$ and the mean square error (MSE) for the $m^{th}$ element of $\x$ is equivalent to the $m^{th}$ diagonal element of the covariance, given by
\begin{equation}
    \text{MSE}(\hat{\x}_m) = E\bigg[\bigg(\hat{\x}_m-\x_m\bigg)^2\bigg] \label{eq:MSE} 
    = E\bigg[\bigg(\hat{\x}_m-E[\x_m]\bigg)^2\bigg] \nonumber 
    = [\bm{C}_{\hat{\x}}]_{mm}.
\end{equation}
Using \eqref{eq:boundElem}, the MSE for the $m^{\rm th}$ element of the unbiased estimator is bounded by the $m^{\rm th}$ element of the inverse of Fisher information matrix, which is substituted from \eqref{eq:GaussianFisher}:
\begin{equation}\label{eq:MSEelement}
    \text{MSE}(\hat{\x}_m) \geq [\bm{I}^{-1}(\x)]_{mm} = \Bigg[ \frac{\partial \h(\x)}{\partial \x_m} \Bigg]^{-1} \cov ~~\Bigg[ \frac{\partial \h(\x)}{\partial \x_m} \Bigg]^{-T}. 
\end{equation}
An estimator with MSE that achieves the equality in \eqref{eq:MSEelement} is referred as an efficient estimator.  

In the rest of this paper, the root-MSE (RMSE) will be used instead of the MSE where $\text{RMSE}(\hat{\x}_m) = \sqrt{\text{MSE}(\hat{\x}_m)}$ so that the RMSE of a UE position estimator is in meters (m).  From \eqref{eq:MSEelement} the RMSE of the two-dimensional estimate of the UE coordinates ($\text{RMSE}_{\text{est}}(\x)$) are lower-bounded by $\text{RMSE}_{\text{CRB}}(\x)$,
where 
\begin{equation}\label{eq:RMSEest}
    \text{RMSE}_{\text{est}}(\x) = \sqrt{E[(\hat{\x}_1-\x_1)^2] + E[(\hat{\x}_2-\x_2)^2]},
\end{equation}
is the RMSE for the unbiased estimator $\hat{\x}$ and 
\begin{equation}\label{eq:RMSE_CRLB}
% \text{RMSE}_{\text{CRLB}}(p_x,p_y) = \sqrt{ \bm{I}_{1,1}(\bm{\theta}) + \bm{I}_{2,2}(\bm{\theta}) }.
\text{RMSE}_{\text{CRB}}(\x) = \sqrt{ \bm{I}_{1,1}^{-1}(\bm{\theta}) + \bm{I}_{2,2}^{-1}(\bm{\theta}) }
\end{equation}
is the CRB.  The UE coordinate estimator RMSE in \eqref{eq:RMSEest} cannot be evaluated exactly, but it can be approximated with Monte-Carlo (MC) simulation,
\begin{equation}\label{eq:RMSE_estSim}\small
\text{RMSE}_{\text{est}}(\x) = \sqrt{ \sum_{\text{i}=1}^{N_{\text{sim}}} \Bigg( \frac{(p_x-\hat{p}_{x_{i}})^2}{N_{\text{sim}}} + \frac{(p_y-\hat{p}_{y_{i}})^2}{N_{\text{sim}}} \Bigg) },
\end{equation} 
where $N_{\text{sim}}$ is the number of Monte-Carlo simulations and $\hat{p}_{x_{i}}$ and $\hat{p}_{y_{i}}$ are the $i^{th}$ Monte-Carlo simulation estimates of the UE position coordinates $p_x$ and $p_y$.

Another bound that was considered was the periodic CRB (PCRB) \cite{routtenberg2011periodic}.  The likelihood function \eqref{eq:likelihood} prevents noisy angular measurements differing from the actual angle by more than $\pi$.  This is a cyclic Gaussian process where very noisy measurements can wrap around and be closer to the actual estimate, which needs to be taken into account in bounding the parameters.  The PCRB bounds such processes and was implemented.  However, it was found that the angular noise required for the PCRB to take effect was larger than what we consider.

\section{Numerical Results and Discussion}\label{section:simulations}

In this section, we evaluate the localization performance of mmWave networks in urban environments.  First, Monte-Carlo simulations of the GAPF estimator and the CRB are used to analyze localization performance as a function of beamwidth. % and the cost of designing networks to meet specific localization metrics is considered. 
Then, mmWave localization performance is studied in urban canyon and urban corner scenarios with one and two FEs using both LOS and NLOS observations, where the urban canyon is represented by two parallel walls and an urban corner is represented by two intersecting orthogonal walls.

\subsection{Localization Performance as a Function of Beamwidth}

\begin{figure}[t]
	\centering
	\subcaptionbox{}{\includegraphics[width=0.21\textwidth]{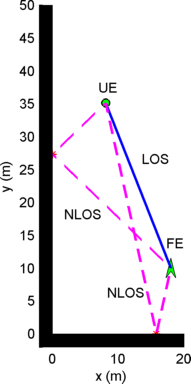}}%
	%\subcaptionbox{}{\includegraphics[width=0.26\textwidth]{CRLB}}%
	\subcaptionbox{}{\includegraphics[width=0.79\textwidth]{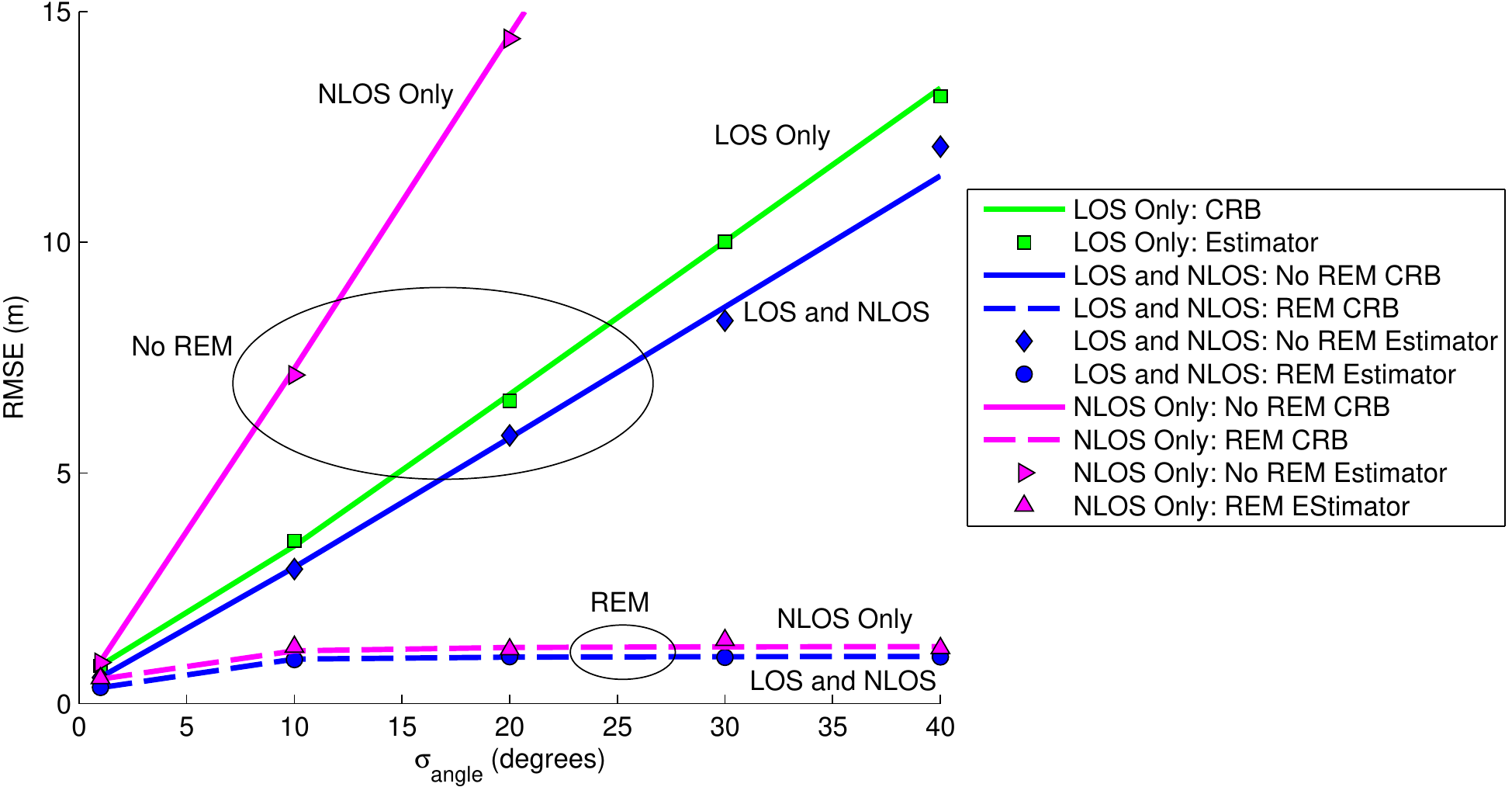}}%
	\caption{(a) Urban corner with one FE, one LOS path, and two NLOS paths. (b) RMSE curves for all path combinations with and without REM for increasing beamwidth.}
	\label{fig:CRBcompareGeom}
\end{figure}

The tradeoff between beamwidth and localization performance is examined in an urban corner scenario with one LOS and two NLOS paths available as seen in Fig.~\ref{fig:CRBcompareGeom}(a).  The urban corner is simulated with a vertical building wall at $x=0$ (m), a horizontal building wall at $y=0$ (m), an FE at $(18,10)$~(m), and a UE at $(8,35)$~(m).  A Monte-Carlo simulation is run where the TOA distance noise standard deviation is fixed at $\sigma_{d_L} = \sigma_{d_N} = 0.75$ (m).  The AOD and AOA noise standard deviations are assumed equal to $\sigma_{\text{angle}}=\sigma_{\alpha_L}=\sigma_{\beta_L}=\sigma_{\alpha_N}=\sigma_{\beta_N}$ and increased.  Equating all of the angular noise standard deviations and varying $\sigma_{\text{angle}}$ is similar to what occurs if a mmWave antenna varies its beamwidth \cite{chethan2018}.  A larger beamwidth results in a larger AOD variance since the it is more likely the path did not travel from the center of the beam, which is what is transmitted as the AOD estimate.  Larger beamwidth also results in more multipath energy, which corrupts the AOA estimate at the receiver and increases the AOA variance. Thus, varying $\sigma_{\text{angle}}$ has similar effects as varying beamwidth and is useful in providing insight into beamwidth effects on localization performance.

% A particular beamwidth can not be linked to specific angular standard deviations because the angular noise is random and depends on the particular urban scenario and propagation environment.  However, increases in beamwidth can be closely represented by increases in $\sigma_{\text{angle}}$, which can be used to analyze performance trade-offs. 

Fig.~\ref{fig:CRBcompareGeom}(b) plots RMSE bounds from the CRB \eqref{eq:RMSE_CRLB} in addition to RMSE from Monte-Carlo simulations \eqref{eq:RMSE_estSim} with the GAPF estimator.  The RMSE is shown for both REM and non-REM systems utilizing different subsets of the available paths from Fig.  \ref{fig:CRBcompareGeom}(a).  Monte-Carlo simulations show the estimator to be closely aligned with the CRB, providing evidence that the GAPF estimator is an efficient estimator. It should be noted that implementation of the particle filter or the gradient method individually leads to RMSE results that are not near the CRB.  This results from the convergence of the estimator to local maxima, and leads to RMSE values that are larger than those given by the CRB.  The individual methods only approach the CRB under certain conditions if initialization is able to provide initial estimates in the immediate vicinity of the global maximum, which is not normally expected.
%\new{It should be noted that implementation of the particle filter or gradient method individually leads to RMSE curves that are not near the CRB.  This results from convergence to local maxima and leads to RMSE values above the CRB.  The individual methods only approach the CRB under certain conditions if initialization is able to provide initial estimates in the immediate vicinity of the global maximum, which is not normally expected.}

For systems without REM, it is observed that increased beam directionality leads to linear improvement in RMSE values and localization performance.  On the other hand, more directionality comes at a cost because it is achieved by adding antenna elements at the transmitter.  It may not be feasible to fit the required number of antenna elements at the transmitter to reach a desired directionality.  Furthermore, the cost of the transmitter grows with each additional antenna element.  Thus, transmitter size constraints and affordability limit the achievable directionality.  Additionally, it is seen that utilizing only the LOS path has much better localization performance than using only the NLOS paths while the addition of NLOS paths to the LOS path only provides modest improvements.  
The REM system RMSE curves reach a threshold where the RMSE does not increase with larger beamwidths.  This occurs because REM provides knowledge of the scatterer locations.  At smaller beamwidths, angles provide precise information and reduce the area of likely UE positions.  Eventually, as the beamwidths increase, a threshold is reached where AOD/AOA measurements become less relevant and the angular uncertainty allows the spread of likely UE positions that are too disparate from the TOA measured path distances.   

This threshold has important implications in designing mmWave systems for localization.  High accuracy localization can be achieved by two methods.  Highly directional beams from antennas with many elements can be implemented, but this can be costly and increases the difficulty of beam alignment.  On the other hand, systems can use REM, which has a high computational cost.   
From this example it is apparent that environments that are mainly LOS, such as rural areas must rely on increasing beam directionality to improve localization performance.  Scenarios with many scatterers and NLOS paths, such as dense urban environments can most easily improve localization performance with the use of REM if the network can handle the computational load. It should be noted that these calculations are under the assumption of perfect REM where the scatterer for each NLOS path is known without any estimation error.  The plotted results are a bound on the performance of any REM system.  Realistic localization with REM will have nonzero errors in scatterer location errors and has many challenges that must be addressed before obtaining small error values.  Thus, an actual REM system will have RMSE values between the non-REM CRB and REM CRB with values depending on the scatterer location estimation error that the REM localization system is able to achieve. 

\subsection{Localization in Urban Environments}
In this section, we consider mmWave localization scenarios in urban canyon and urban corner environments where 5G mmWave cellular networks are expected to first get deployed and there may be one or two FEs available to be used for localization purposes.  The performance with/without the use of REM for localization is analyzed, where with REM, we assume the scatterer locations are available as in \eqref{eq:paramSpaceREM}. Frequencies of 28 GHz and 73 GHz are considered where Table~\ref{table:param} is used to define simulation parameters.  Since the GAPF estimator is closely aligned with the CRB, system performance is determined by evaluation of the CRB~\eqref{eq:RMSE_CRLB} for each scenario to obtain RMSE. The results for each scenario are shown in identically organized figures where (a) shows the geometry under consideration for an example UE location, (b) plots the RMSE cumulative distribution function (CDF) curve for all possible UE locations with and without REM for 28 GHz and 73 GHz systems, (c) is a contour plot of RMSE values ($\text{RMSE}_{\text{NoREM}}$) for UE locations throughout the scenario without REM at 73 GHz, (d) is a contour plot of RMSE values ($\text{RMSE}_{\text{REM}}$) for UE locations throughout the scenario utilizing REM at 73 GHz,  and (e) is a contour plot of the improvement in RMSE from non-REM to REM at 73 GHz where $\Delta\text{RMSE} = \text{RMSE}_{\text{NoREM}} - \text{RMSE}_{\text{REM}}$.

\subsubsection{Urban Canyon}

\begin{figure}[t]
	\centering
	\subcaptionbox{}{\includegraphics[width=0.20\textwidth]{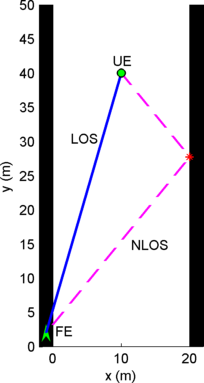}}%
	%\subcaptionbox{}{\includegraphics[width=0.26\textwidth]{CRLB}}%
	%\hfill
	\subcaptionbox{}{\includegraphics[width=0.47\textwidth]{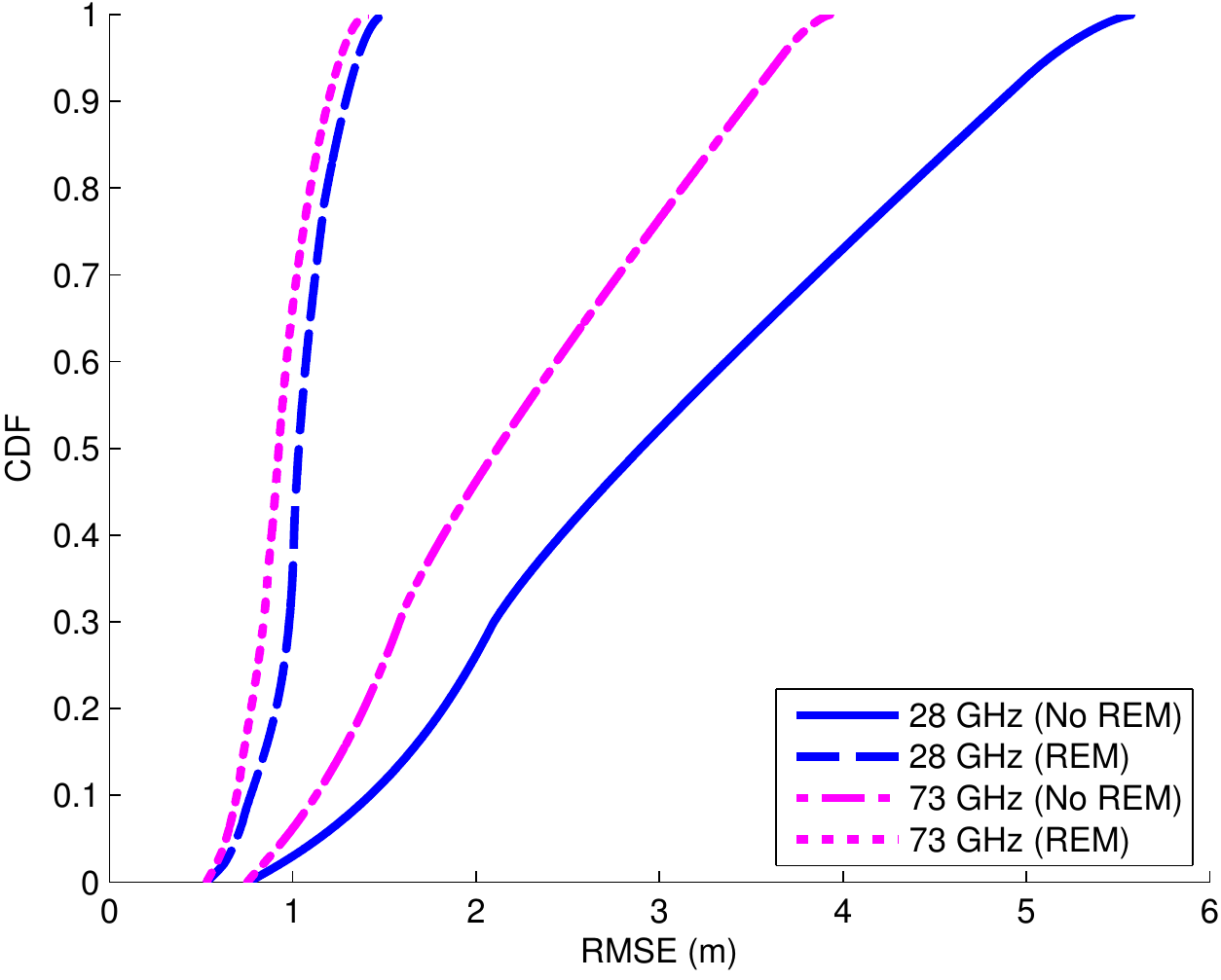}}%

    \subcaptionbox{}{\includegraphics[width=0.25\textwidth]{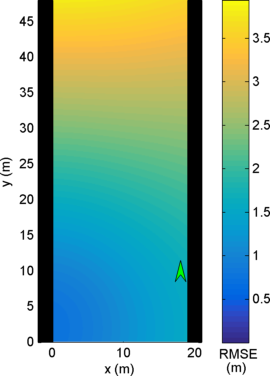}}%
	%\subcaptionbox{}{\includegraphics[width=0.26\textwidth]{CRLB}}%
	%\hfill
	\subcaptionbox{}{\includegraphics[width=0.25\textwidth]{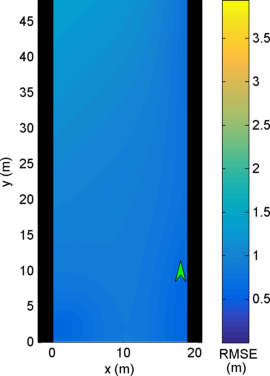}}%
	\subcaptionbox{}{\includegraphics[width=0.25\textwidth]{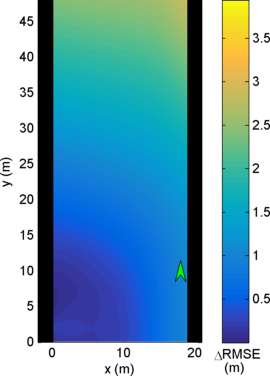}}%
	%\vspace{-0.5cm}
	\hfill
	\caption{(a) Urban canyon with one FE, one LOS path, and one NLOS path.  (b) CDF of RMSE. (c) RMSE at 73 GHz without REM. (d) RMSE at 73 GHz with REM. (e) Improvement in RMSE with REM at 73 GHz.}
	\label{fig:parallel1}
\end{figure}

An urban canyon is simulated with parallel walls at $x = 0$ (m) and $ x = 20$ (m) as shown in Fig. \ref{fig:parallel1}(a) with an example set of paths for an FE on the left wall at $(-1,2)$ (m) and a UE located at $(10,40)$ (m).  In this case, a single LOS and a single NLOS path reflected by the right wall are received by the UE from the FE.  From Fig. \ref{fig:parallel1}(b), it is seen that localization performance is improved with the addition of REM as expected.  As with all scenarios, 73 GHz performs slightly better than 28 GHz.  Referring to Table~\ref{table:param}, 73 GHz is equivalent to 28 GHz in timing estimation, but has better estimation of AOA and AOD.  As it is at a higher frequency, more antennas can be fit on a chip, allowing a more directional beam.  The more directional beam captures less multipath and reduces the variance on AOA and AOD estimation, enabling improved performance.  Figs. \ref{fig:parallel1}(c)-(e) show that REM provides little improvement when the UE is near the FE, but significant improvement far from the FE.  As with every scenario, REM performs better closer to the reflecting walls.  This can be understood by treating the scatterers as additional FEs, which are known as a result of REM.  Thus, UE locations near the wall give lower RMSE because they are closer to the these FEs for each path.

\begin{figure}[t]
	\centering
	\subcaptionbox{}{\includegraphics[width=0.2\textwidth]{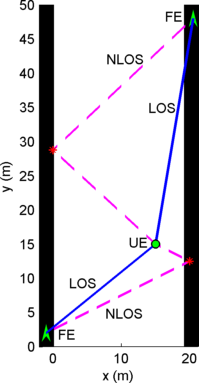}}%
	%\subcaptionbox{}{\includegraphics[width=0.26\textwidth]{parallelCRLB_73GHz_2BS_2LOS2NLOSpathPlot}}%
	%\hfill
	\subcaptionbox{}{\includegraphics[width=0.47\textwidth]{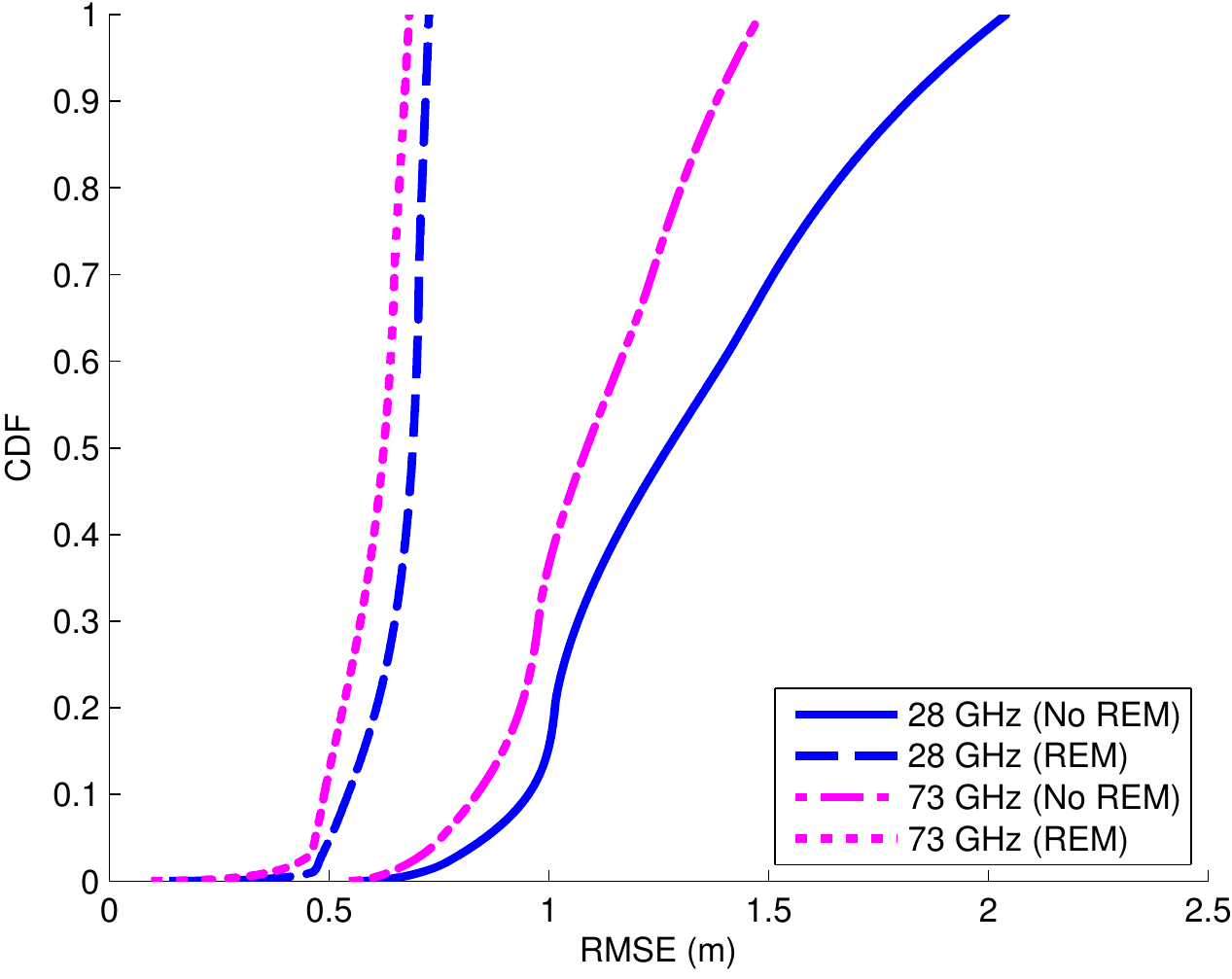}}%
	
	\subcaptionbox{}{\includegraphics[width=0.25\textwidth]{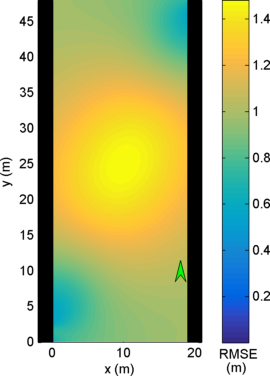}}%
	%\subcaptionbox{}{\includegraphics[width=0.26\textwidth]{CRLB}}%
	%\hfill
	\subcaptionbox{}{\includegraphics[width=0.25\textwidth]{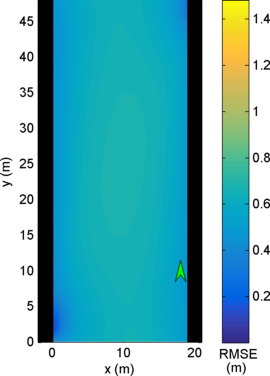}}%
	\subcaptionbox{}{\includegraphics[width=0.25\textwidth]{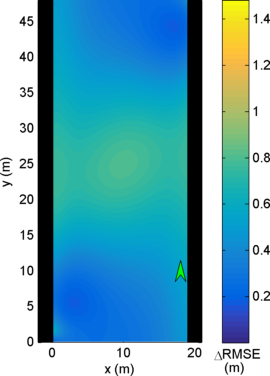}}%
	%\vspace{-0.5cm}
	\hfill
	\caption{(a) Urban canyon with two FEs, two LOS paths, and two NLOS paths. (b) CDF of RMSE. (c) RMSE at 73 GHz without REM. (d) RMSE at 73 GHz with REM. (e) Improvement in RMSE with REM at 73 GHz.}
	\label{fig:parallel2}
\end{figure}

Another scenario adds a second FE to the urban canyon at $(21,48)$ (m), which has an additional LOS link and a NLOS link reflected from the left wall as seen in Fig. \ref{fig:parallel2}(a).  The RMSE curve seen in Fig.~\ref{fig:parallel2}(b) shows that REM again greatly improves localization performance.  We also realize from Fig.~\ref{fig:parallel2}(b) that adding a second mmWave BS significantly improves the localization accuracy with and without REM.  Figs. \ref{fig:parallel2}(c)-(e) show that REM offers more performance gain as the UE travels further from either FE.  

\subsubsection{Urban Corner}

\begin{figure}[t]
	\centering
	\subcaptionbox{}{\includegraphics[width=0.19\textwidth]{corner1LOS2NLOS.png}}%
	%\hfill
	%\subcaptionbox{}{\includegraphics[width=0.26\textwidth]{orthogonalCRLB_73GHz_1BS_1LOS2NLOSpathPlot}}%
	%\hfill
	\subcaptionbox{}{\includegraphics[width=0.47\textwidth]{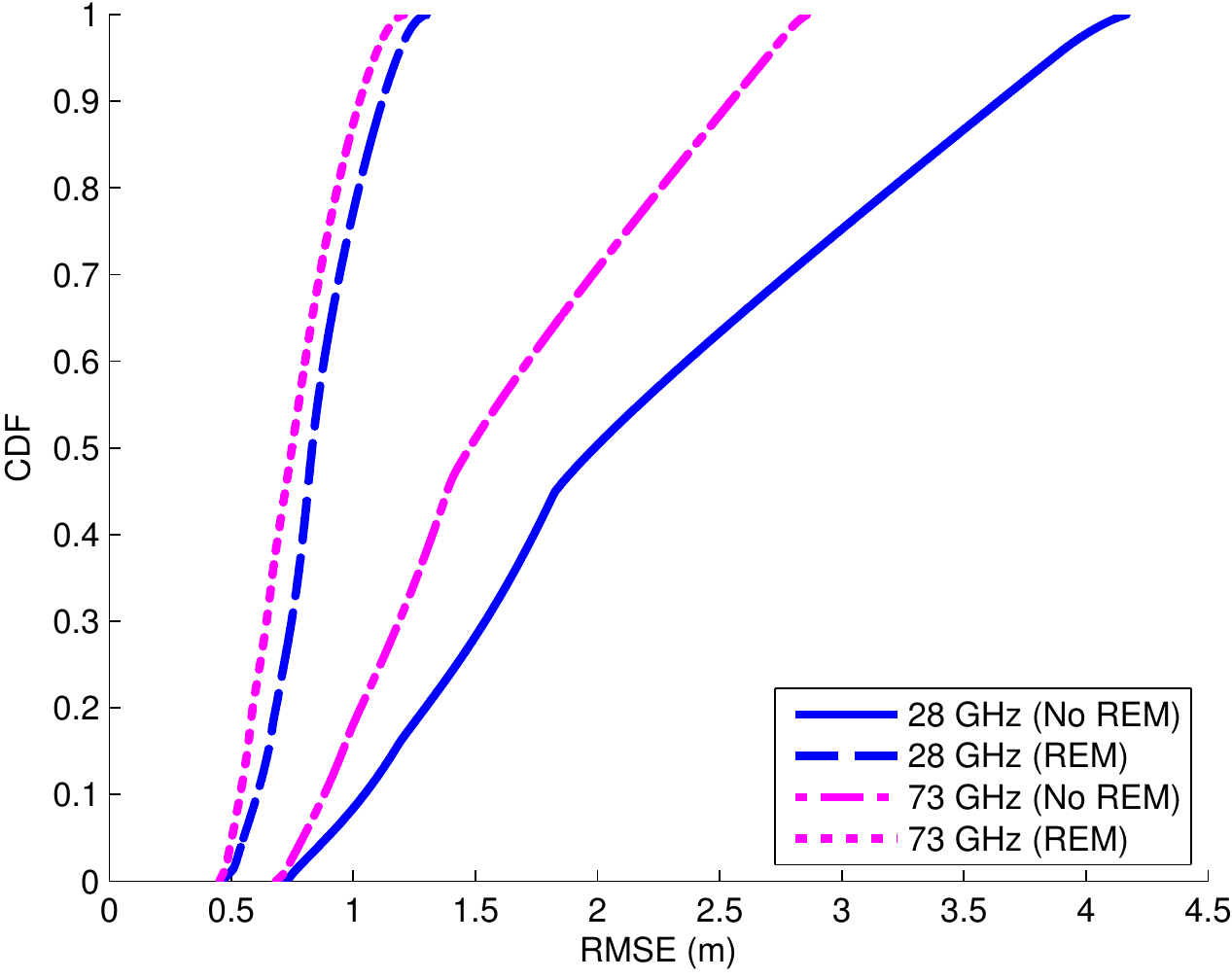}}%
	
	\subcaptionbox{}{\includegraphics[width=0.25\textwidth]{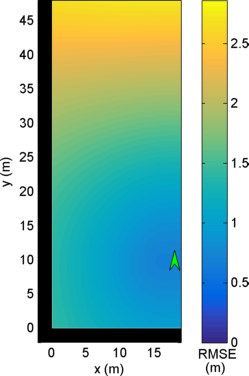}}%
	%\subcaptionbox{}{\includegraphics[width=0.26\textwidth]{CRLB}}%
	%\hfill
	\subcaptionbox{}{\includegraphics[width=0.25\textwidth]{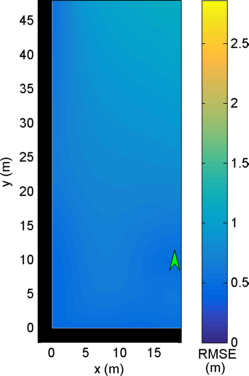}}%
	\subcaptionbox{}{\includegraphics[width=0.25\textwidth]{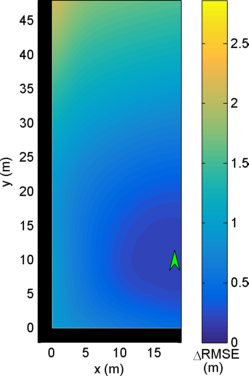}}%
	%\vspace{-0.5cm}
	\hfill
	\caption{(a) Urban corner with one FE, one LOS path, and two NLOS paths.  (b) CDF of RMSE. (c) RMSE at 73 GHz without REM. (d) RMSE at 73 GHz with REM. (e) Improvement in RMSE with REM at 73 GHz.}
	\label{fig:corner1}
\end{figure}

An urban corner is simulated with a vertical building wall at $x=0$ (m) and a horizontal building wall at $y=0$ (m) as seen in Fig.~\ref{fig:corner1}(a) with an example set of paths for a FE at $(18,10)$ (m) and a UE at $(15,15)$ (m).  In this case, two LOS paths and two NLOS paths, one reflected from each wall are received by the UE from the FE. Fig. \ref{fig:corner1}(b) shows that the geometry of the corner allows better localization accuracy relative to the canyon as it allows more NLOS paths.  From Figs. \ref{fig:corner1}(c)-(e) it is seen that REM provides the most gain far from the FE, which is most significant close to the left reflected wall.  

\begin{figure}[t]
	\centering
	\subcaptionbox{}{\includegraphics[width=0.2\textwidth]{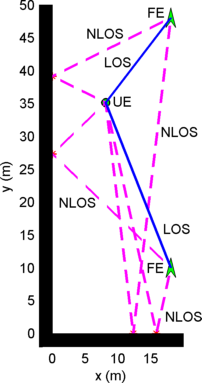}}%
	%\hfill
	%\subcaptionbox{}{\includegraphics[width=0.26\textwidth]{orthogonalCRLB_73GHz_2BS_2LOS4NLOSpathPlot}}%
	%\hfill
	\subcaptionbox{}{\includegraphics[width=0.47\textwidth]{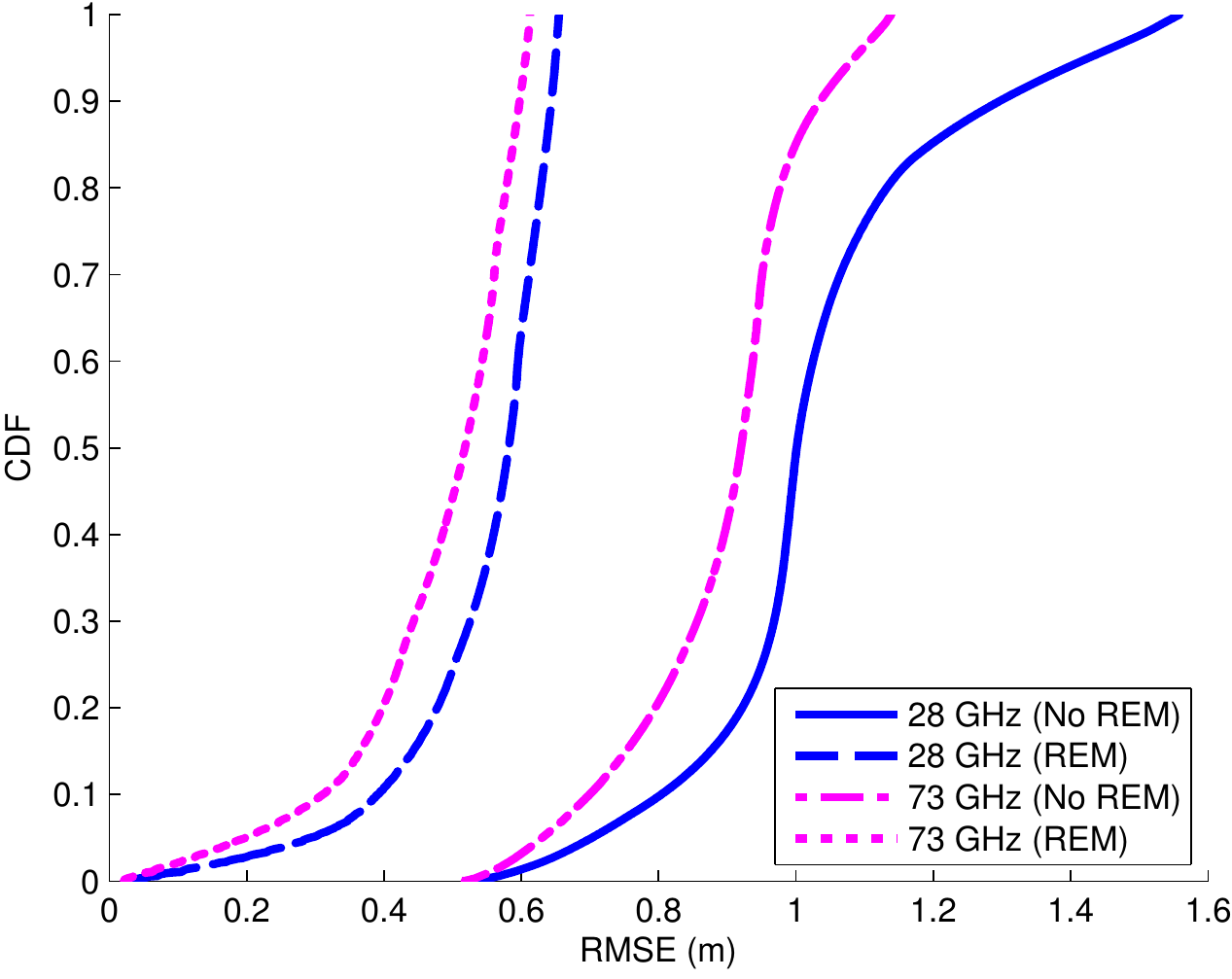}}%
	
	\subcaptionbox{}{\includegraphics[width=0.25\textwidth]{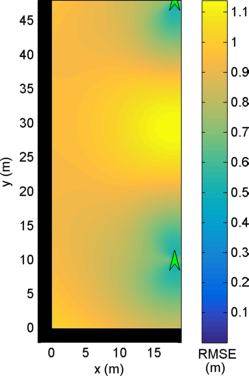}}%
	%\subcaptionbox{}{\includegraphics[width=0.26\textwidth]{CRLB}}%
	%\hfill
	\subcaptionbox{}{\includegraphics[width=0.25\textwidth]{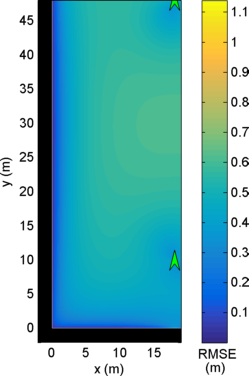}}%
	\subcaptionbox{}{\includegraphics[width=0.25\textwidth]{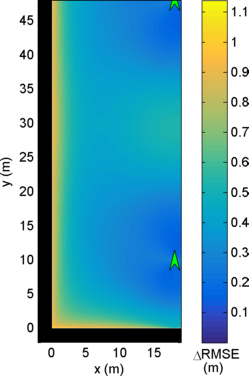}}%
	%\vspace{-0.5cm}
	\hfill
	\caption{(a) Urban corner with two FEs, two LOS paths, and two NLOS paths.  (b) CDF of RMSE. (c) RMSE at 73 GHz without REM. (d) RMSE at 73 GHz with REM. (e) Improvement in RMSE with REM at 73 GHz.}
	\label{fig:corner2}
\end{figure} 

A second FE can be added to the urban corner scenario at $(18,48)$ (m) as in Fig.~\ref{fig:corner2}(a). The geometry of the corner allows four NLOS paths in addition to two LOS paths, providing large amounts of information for localization.  It is seen from Fig.~\ref{fig:corner2}(b) that the use of many paths results in high localization accuracy. Figs.~\ref{fig:corner2}(c)-(e) show that localization performance does not only depend on the distance from the FEs.  The midpoint between the two FEs has worse performance than points further from both FEs that are closer to the left wall.  

This scenario is similar to what is expected for the first deployed 5G networks where there may be multiple FE in range and many LOS/NLOS links may be available and exploited for significant gains in localization performance.  These results provide evidence that 5G networks using two-step localization in environments rich in available links can expect sub-meter localization accuracy even without REM.

\subsection{Building REM from Localization Outcomes}

\begin{figure}[!ht]
	\centering
	\subcaptionbox{}{\includegraphics[width=0.26\textwidth]{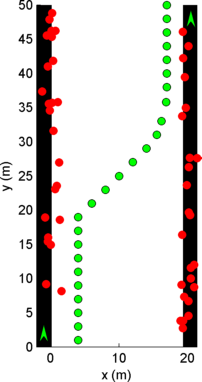}}%
	%\subcaptionbox{}{\includegraphics[width=0.26\textwidth]{orthogonalCRLB_73GHz_2BS_2LOS4NLOSpathPlot}}%
	\hspace{1.0cm}
	\subcaptionbox{}{\includegraphics[width=0.25\textwidth]{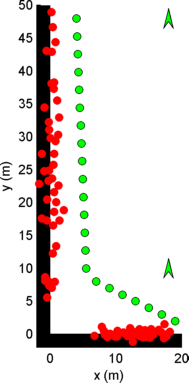}}%
% 	\hspace{5.5cm}
	\hfill
	\caption{Mapped scatterer locations (red) generated from GAPF estimator at UE locations (green) on a path through (a) the urban canyon and (b) the urban corner.}
	\label{fig:SLAM}
\end{figure} 

The scatterer locations are nuisance parameters in the estimation of the UE position.  However, the scatterer locations are valuable information and can be used to create an REM. Fig.~\ref{fig:SLAM} plots the estimated scatterer locations (red) obtained from NLOS paths for a given UE trajectory (green dots).  Fig.~\ref{fig:SLAM}(a) shows the estimated scatterer locations from the urban canyon in the scenario shown in Fig.~\ref{fig:parallel2}.  It is seen that scatterer estimates are along the entire length of both the left and right building as both of the FEs are able to establish paths that reflect from both buildings, providing good coverage of the environment. Fig.~\ref{fig:SLAM}(b) shows the estimated scatterer locations from the urban corner in the scenario shown in Fig.~\ref{fig:corner2}.  In this scenario, each FE is able to establish two NLOS paths for each UE location.  However, it is seen that the very interior of the corner has no scatterer coverage.  Coverage is possible if a UE travels further into the corner, but it should be noted that REM may not always provide complete coverage of the environment. The scatterer locations can be stored with the AOD/AOA/TOA parameters associated with them to create an REM.  An algorithm that uses the information from REM to assist in localization is left for future work.

\section{Conclusion}\label{sec:conclusion}

In this paper, we analyze the localization performance of a reduced complexity method for 5G mmWave urban networks with multiple available LOS/NLOS links with one or more FEs.  %The first step of the two-step approach separately estimates AOA, AOD, and TOA for each path.  Then, the second step estimates the scatterer locations for NLOS paths and estimates the UE location.  
Specifically, we study the localization performance in urban canyon and urban corner settings utilizing AOA, AOD, and TOA measurements at a UE from one or more FEs. We consider scenarios with and without REM, where in the latter case, all the scatterer locations for NLOS links are also simultaneously estimated along with the UE location. This results in a high dimensional unknown parameter space. As a consequence, mmWave localization requires processing of a likelihood function for the unknown parameter vector with many local maxima, making it difficult to find the global optimum solution. To address this problem, we propose a GAPF estimator which combines particle filter and gradient method, and Monte-Carlo simulations show that its  performance matches that of the CRB in terms of localization RMSE. The scatterer locations that are estimated using the proposed algorithm can be used to create an REM for an urban environment.

Our results show that when REM is used with the proposed two-step approach, sub-meter localization accuracy is feasible in mmWave urban networks using even a single FE. However, further research is required for evaluating the feasibility of realizing REM effectively. Without REM (when scatterer locations need to be estimated simultaneously), median RMSE lower than two meters is possible with a single FE, and lower than a meter is possible with two FEs. Results show that the urban corner provides better localization performance due to availability of larger number of NLOS paths.  Thus, dense urban environments with non-trivial building structures and many scatterers are best suited for mmWave localization.  It is shown that localization is improved by increasing the number of antenna elements to increase beam directionality or by implementing REM to assist in estimating NLOS scatterer locations, both of which come with the trade-offs of higher cost and complexity.

\section*{Appendix: Elements of the Jacobian Matrix}
\label{Sec:appendixa}

The Jacobian in \eqref{eq:Jacobian} is calculated as:
\begin{align}\label{Eq:Appdx1}
    \frac{\partial \h(\x)}{\partial \x} = 
    \begin{bmatrix}
    \frac{\partial \bm{\alpha}(\x)}{\partial p_x} & \frac{\partial \bm{\alpha}(\x)}{\partial p_y} & \frac{\partial \bm{\alpha}(\x)}{\partial s_x(1)} & \cdots & \frac{\partial \bm{\alpha}(\x)}{\partial s_x(N_N)} & \frac{\partial \bm{\alpha}(\x)}{\partial s_y(1)} & \cdots & \frac{\partial \bm{\beta}(\x)}{\partial s_y(N_N)} \\
    \frac{\partial \bm{\beta}(\x)}{\partial p_x} & \frac{\partial \bm{\beta}(\x)}{\partial p_y} & \frac{\partial \bm{\beta}(\x)}{\partial s_x(1)} & \cdots & \frac{\partial \bm{\beta}(\x)}{\partial s_x(N_N)} & \frac{\partial \bm{\beta}(\x)}{\partial s_y(1)} & \cdots & \frac{\partial \bm{\beta}(\x)}{\partial s_y(N_N)} \\
    \frac{\partial \bm{d}(\x)}{\partial p_x} & \frac{\partial \bm{d}(\x)}{\partial p_y} & \frac{\partial \bm{d}(\x)}{\partial s_x(1)} & \cdots & \frac{\partial \bm{d}(\x)}{\partial s_x(N_N)} & \frac{\partial \bm{d}(\x)}{\partial s_y(1)} & \cdots & \frac{\partial \bm{d}(\x)}{\partial s_y(N_N)}
    \end{bmatrix},
\end{align}
and the corresponding derivatives in~\eqref{Eq:Appdx1} for the AOA, AOD, and distance with respect to \emph{UE} coordinates considering LOS and NLOS links are given by:
\begin{align}
    \frac{\partial \alpha_{j,k_j,0}(\x)}{\partial p_x} &= \frac{p_y-q_y(k_j)}{(p_x-q_x(k_j))^2 + (p_y-q_y(k_j))^2}, \\ %\quad %\\
    \frac{\partial \alpha_{j,k_j,0}(\x)}{\partial p_y} &= \frac{q_x(k_j)-p_x}{(p_x-q_x(k_j))^2 + (p_y-q_y(k_j))^2}, \\
    \frac{\partial \beta_{j,k_j,0}(\x)}{\partial p_x} &= \frac{p_y-q_x(k_j)}{(p_x-q_x(k_j))^2 + (p_y-q_y(k_j))^2}, \\ %\quad %\\
    \frac{\partial \beta_{j,k_j,0}(\x)}{\partial p_y} &= \frac{q_x(k_j)-p_x}{(p_x-q_x(k_j))^2 + (p_y-q_y(k_j))^2}, \\
    \frac{\partial d_{j,k_j,0}(\x)}{\partial p_x} &= \frac{p_x-q_x(k_j)}{\sqrt{(p_x-q_x(k_j))^2 + (p_y-q_y(k_j))^2}}, \\ %\quad %\\
    \frac{\partial d_{j,k_j,0}(\x)}{\partial p_y} &= \frac{p_y-q_y(k_j)}{\sqrt{(p_x-q_x(k_j))^2 + (p_y-q_y(k_j))^2}}, \\
    \frac{\partial \alpha_{j,k_j,i_j}(\x)}{\partial p_x} &= \frac{p_y-s_y(i_j)}{(s_x(i_j)-p_x)^2 + (s_y(i_j)-p_y)^2}, \\ % \quad %\\
    \frac{\partial \alpha_{j,k_j,i_j}(\x)}{\partial p_y} &= \frac{s_x(i_j)-p_x}{(s_x(i_j)-p_x)^2 + (s_y(i_j)-p_y)^2},\\
    \frac{\partial d_{j,k_j,i_j}(\x)}{\partial p_x} &= \frac{p_x-s_x(i_j)}{\sqrt{(s_x(i_j)-p_x)^2 + (s_y(i_j)-p_y)^2}}, \\ % \quad %\\
    \frac{\partial d_{j,k_j,i_j}(\x)}{\partial p_y} &= \frac{p_y-s_y(i_j)}{\sqrt{(s_x(i_j)-p_x)^2 + (s_y(i_j)-p_y)^2}},
\end{align}
where the derivatives of the AOD with respect to UE coordinates for the NLOS case are assumed zero based on~\eqref{eq:parameters5}. Note that if the wall orientation is available, it can be possible to relate AOD explicitly to UE location for the urban localization setting, which is left as a future work. 

The derivatives in~\eqref{Eq:Appdx1} for the AOA, AOD, and distance with respect to \emph{scatterer} coordinates considering NLOS links are given by: \begin{align} 
    \frac{\partial \alpha_{j,k_j,i_j}(\x)}{\partial s_x(i_j)} &= \frac{s_y(i_j)-p_y}{(s_x(i_j)-p_x)^2 + (s_y(i_j)-p_y)^2}, \\
    \frac{\partial \alpha_{j,k_j,i_j}(\x)}{\partial s_y(i_j)} &= \frac{p_x-s_x(i_j)}{(s_x(i_j)-p_x)^2 + (s_y(i_j)-p_y)^2}, \\
    \frac{\partial \beta_{j,k_j,i_j}(\x)}{\partial s_x(i_j)} &= \frac{s_y(i_j)-q_y(k_j)}{(s_x(i_j)-q_x(k_j))^2 + (s_y(i_j)-q_y(k_j))^2}, \\
    \frac{\partial \beta_{j,k_j,i_j}(\x)}{\partial s_y(i_j)} &= \frac{q_x(k_j)-s_x(i_j)}{(s_x(i_j)-q_x(k_j))^2 + (s_y(i_j)-q_y(k_j))^2}, \\
    % \end{align}
    % \begin{align}
    \frac{\partial d_{j,k_j,i_j}(\x)}{\partial s_x(i_j)} &= \frac{s_x(i_j)-p_x}{\sqrt{(s_x(i_j)-p_x)^2 + (s_y(i_j)-p_y)^2}} \nonumber \\
    &+ \frac{s_x(i_j)-q_x(k_j)}{ \sqrt{(s_x(i_j)-q_x(k_j))^2 + (s_y(i_j)-q_y(k_j))^2}}, \\
    \frac{\partial d_{j,k_j,i_j}(\x)}{\partial s_y(i_j)} &= \frac{s_y(i_j)-p_y}{\sqrt{(s_x(i_j)-p_x)^2 + (s_y(i_j)-p_y)^2}} \nonumber \\ 
    &+ \frac{s_y(i_j)-q_y(k_j)}{ \sqrt{(s_x(i_j)-q_x(k_j))^2 + (s_y(i_j)-q_y(k_j))^2}},
\end{align}
and all other remaining elements are zero since there are no scatterers for LOS case.

\section*{Abbreviations}
\begin{table}[H]
    \centering
    \begin{tabular}{l l}
         \textbf{mmWave:} &  Millimeter Wave  \\
         \textbf{GPS:} & Global positioning system \\
         \textbf{LOS:} & Line-of-sight \\
         \textbf{LOS:} & Line-of-sight \\
         \textbf{NLOS:} & Non-line-of-sight \\
        \textbf{TOA:} & Time-of-arrival \\
        \textbf{AOD:} & Angle-of-departure \\
        \textbf{AOA:} & Angle-of-arrival \\
        \textbf{TDOA:} & time-difference-of-arrival \\
        \textbf{UE:} & User equipment \\
        \textbf{FE:} & Fixed equipment \\
        \textbf{GAPF:} & Gradient-assisted particle filter \\
        \textbf{REM:} & Radio-environmental mapping \\
        \textbf{ML:} & Maximum  likelihood \\
        \textbf{CRB:} & Cramer-Rao bound \\
        \textbf{PCRB:} & Periodic Cramer-Rao bound \\
        \textbf{MSE:} & Mean square error \\
        \textbf{RMSE:} & Root-mean square error \\
        \textbf{CDF:} & Cumulative distribution function \\
    \end{tabular}
    \caption*{}
    \label{table:Abbreviations}
\end{table}

\section*{Declarations:}

\section*{Availability of data and material}
  The datasets supporting the conclusions of this article are included within the article.

\section*{Competing interests}
  The authors declare that they have no competing interests.

\section*{Funding}

This work has been supported in part by the National Aeronautics and Space Administration under the Federal Award ID number NNX17AJ94A

\section*{Author's contributions}
    Ideas are developed collaboratively among the two authors. All simulations are carried out by the first author, and improvements are integrated after discussions among the authors. Paper has been written collaboratively by the two authors with major effort coming from the first author.
    
% \section*{Acknowledgements}
%     None

%%  After submission of the .TEX file,                     %%
%%  you will be prompted to submit your .BBL file.         %%

% if your bibliography is in bibtex format, use those commands:
\bibliographystyle{bmc-mathphys} % Style BST file (bmc-mathphys, vancouver, spbasic).
\bibliography{mmWaveLocPaperBib}      % Bibliography file (usually '*.bib' )

\end{document}